\documentclass[11pt]{article}

\usepackage{amsmath}
\usepackage{amssymb}
\usepackage{amsthm}
\usepackage{epsfig}
\usepackage{xcolor}
\usepackage{graphicx}
\usepackage{bm}
\usepackage{enumitem}
\usepackage{booktabs}
\usepackage{algpseudocodex}
\usepackage{algorithm}
\usepackage{authblk}
\usepackage{cite}
\usepackage{url}
\usepackage{xcolor}
\usepackage{hyperref}
\usepackage{wrapfig}
\hypersetup{
    colorlinks,
    linkcolor={blue!50!black},
    citecolor={blue!50!black},
}
\colorlet{linkequation}{blue}

\usepackage{caption}
\newtheorem{theorem}{Theorem}
\newtheorem{proposition}{Proposition}

\allowdisplaybreaks
\usepackage[top=1in,bottom=0.95in,left=1in,right=1in]{geometry}
\usepackage[utf8]{inputenc}

\def\reals{\mathbb{R}}

\def\Fit{\textsf{Fit}}
\def\Unfit{\textsf{Unfit}}
\def\EE{\mathbb E}
\def\PP{\mathbb P}
\def\Exp{\mathrm{Exp}}
\def\EventType{\mathrm{EventType}}

\def\de{\mathrm{d}}

\def\cE{\mathcal{E}}

\def\cR{\mathcal{R}}
\def\cT{\mathcal{T}}

\def\<{\langle}
\def\>{\rangle}

\def\FE{\mathsf{FE}}

\renewcommand{\P}{\mathbb{P}} 
\newcommand{\E}{\mathbb{E}} 
\newcommand{\R}{\mathbb{R}} 
\newcommand{\Law}{\mathsf{Law}}

\newcommand{\gbar}{\overline{g}{}}

\newcommand{\tmax}{t_\mathrm{max}}
\newcommand{\nmax}{n_{\max}}

\def\suppfref#1{{Supplementary Figure~\ref{#1}}}
\def\supptext#1{{Appendix~\ref{#1}}}

\usepackage{xspace}
\newcommand{\phenotype}{type\xspace}
\newcommand{\phenotypes}{types\xspace}

\newcommand{\beginsupplement}{%
        \setcounter{table}{0}
        \renewcommand{\thetable}{S\arabic{table}}%
        \setcounter{figure}{0}
        \renewcommand{\thefigure}{S\arabic{figure}}%
        \numberwithin{equation}{section}
     }

\title{Exact and efficient phylodynamic simulation from\break arbitrarily large populations}

\author[1]{Michael Celentano}
\author[2]{William S. DeWitt}
\author[2]{Sebastian Prillo}
\author[1,2]{Yun S. Song}

\affil[1]{Department of Statistics, University of California, Berkeley}
\affil[2]{Computer Science Division, University of California, Berkeley}

\date{}

\begin{document}

\maketitle

\begin{abstract}
Many biological studies involve inferring the evolutionary history of a sample of individuals from a large population and interpreting the reconstructed tree. Such an ascertained tree typically represents only a small part of a comprehensive population tree and is distorted by survivorship and sampling biases. Inferring evolutionary parameters from ascertained trees requires modeling both the underlying population dynamics and the ascertainment process. A crucial component of this phylodynamic modeling involves tree simulation, which is used to benchmark probabilistic inference methods. To simulate an ascertained tree, one must first simulate the full population tree and then prune unobserved lineages. Consequently, the computational cost is determined not by the size of the final simulated tree, but by the size of the population tree in which it is embedded. In most biological scenarios, simulations of the entire population are prohibitively expensive due to computational demands placed on lineages without sampled descendants. Here, we address this challenge by proving that, for any partially ascertained process from a general multi-type birth-death-mutation-sampling model, there exists an equivalent process with \textit{complete sampling} and \textit{no death}, a property which we leverage to develop a highly efficient algorithm for simulating trees. Our algorithm scales linearly with the size of the final simulated tree and is independent of the population size, enabling simulations from extremely large populations beyond the reach of current methods but essential for various biological applications. We anticipate that this unprecedented speedup will significantly advance the development of novel inference methods that require extensive training data.
\end{abstract}

\section{Introduction}

Phylogenetic trees describe the ancestral relationships within a sample of individuals from a large population and are central objects in studies of evolution. (We make no distinction between phylogenetic and genealogical trees in this work.)
These individuals can represent extant or extinct species in macroevolution \cite{Morlon2014,MaddisonMidfordOtto2007},
viral sequences collected over time in epidemiology \cite{stadler2010},
or individual cells in studies of cancer evolution or affinity maturation \cite{QuinnJonesOkimotoEtAl2021,Yang2022,HornsVollmersCrooteEtAl2016,Horton2022,Hoehn2023-hv}, for example.
The branch lengths and branching rates in these phylogenies have been used to infer shifts in diversification rates across lineages and over time in the tree of life \cite{Hohna2011,FitzJohn2009,Louca:2020aa}, to estimate effective reproduction number or predict future strains in viral evolution \cite{Stadler2011,neherRusselShraiman2014},
and to estimate rates of metastasis, phylodynamics, plasticity, or paths of tumor evolution in cancer \cite{2022_Yang_KP},
among many other applications.

Due to death and incomplete sampling in these applications, the inferred phylogeny typically represents a partial history of the full population.
This partial observation process can lead to bias in the estimation of fundamental population parameters.
For example, the branching rate in the observed phylogeny is typically substantially smaller than that in the full population phylogeny \cite{NeeMayHarvey1994,MooreHohnaMayRannalaHuelsenbeck2016}.
Moreover, certain population-level quantities may not be identified by the distribution of the observed phylogeny without further assumptions \cite{Louca:2020aa,legried2022class,legried2023identifiability}.

Simulations play a central role in tree-based inference.
Their uses include benchmarking existing methods \cite{TitleRabosky2019,QuinnJonesOkimotoEtAl2021},
training novel methods based on simulated data \cite{ThompsonLiebeskindScullyLandis2023},
and, in some approaches, may form a sub-routine of the inferential method itself \cite{XieValentaEtienne2023}.
In order to faithfully represent the relationship between population-level parameters and the observed data,
these simulations must include the partial observation process.
Current approaches simulate the partial observation process directly.
First, the phylogeny of the full population is simulated;
next, some lineages are sampled;
and finally, lineages that are not ancestral to a sampled lineage are pruned from the full phylogeny \cite{NeeMayHarvey1994}.

In many applications,
full-population simulations are prohibitively expensive because they must expend substantial computational resources on lineages that, because they have no sampled descendants, do not appear in the observed phylogeny.
For example, in a typical year, there are tens of millions of flu infections in the United States alone \cite{Rolfes2018-yz}.
Nevertheless, large phylogenetic analyses of flu are often limited to tens of thousands of sampled viral sequences \cite{barratCharlaixEtAl2021},
and it is not uncommon to analyze phylogenies with only a few hundred samples per year \cite{NeherBedford2015}.
In cancer, a cubic centimeter of tumor mass contains roughly one billion cells, yet single-cell resolution studies of cancer with CRISPR-Cas9-based lineage tracing assays sequence fewer than 100,000 cells \cite{2022_Yang_KP, 2021_Xenograft_Quinn_Science, SIMEONOV20211150, 2020_Cassiopeia_Jones}, representing roughly $0.01\%$ of the population.
Recent works at best simulate a population of $40,000$ cancer cells subsequently sub-sampled down to $1\%$ \cite{prillo_convexml:_2023, 2020_Cassiopeia_Jones}, which is highly unrealistic.

This paper proposes a novel algorithm for exact (up to numerical and time-discretization error) simulation from a general class of multi-type birth-death processes with birth, death, mutation, and incomplete sampling.
This class of models agrees with those considered by MacPherson et al.~\cite{macPhersonLoucaMcLauglinJoyPennell2021},
which unified a collection of models used in a wide range of biological contexts, including the applications described above \cite{Hohna2011,Louca:2020aa,Stadler2011,FitzJohn2009,Morlon2014,LaudannoHaegemanRaboskyEtienne2020,MorlonParsonsPlotkin2011,MaddisonMidfordOtto2007,neherHallatschek2013}. 
This general class of models is referred to as ``birth-death-mutation-sampling'' (BDMS) models.
The computational cost of our algorithm for simulating BDMS models scales not with the size of the full population, as do existing approaches, but rather with the sample size.
It thus avoids expending computational resources on unobserved lineages and is essentially optimal in terms of runtime. 
Based on the typical numbers for flu and cancer evolution studies described above,
our approach can reduce computation by a factor of 1,000 to 10,000,
and make it feasible to simulate with realistic population-size and sub-sampling parameters.
Thus, our novel algorithm will facilitate the assessment of existing methods under more biologically realistic settings and enable the development of simulation-based training and estimation approaches.

From a technical point of view,
our algorithm is based on the following insight.
The distribution of the observed phylogeny in any BDMS model is equivalent to that generated by an alternative BDMS model with no death and complete sampling.
In this alternative BDMS model, which we call the \emph{forward-equivalent model}, the full phylogeny corresponds to the observed phylogeny.
Therefore, simulating from the forward-equivalent model allows us to avoid expending computational resources on unobserved lineages while generating observed trees from the same distribution.

Our insight is closely related to recent work on statistical identifiability in sub-sampled phylogenetic birth-death models \cite{Louca:2020aa,legried2022class,legried2023identifiability}.
This work observed that in the single-type setting,
different time-dependent population-level parameterizations can give rise to the same observed data distribution.
Our algorithm is based on the lack of identifiability in the more general multi-type setting.
Whereas lack of identifiability prohibits exact inference (without further assumptions),
it also facilitates simulation.
Because multiple BMDS models give rise to the same observed data distribution, 
we can base our simulations on those that can be simulated most efficiently.

\section{The multi-type birth-death-mutation-sampling (BDMS) model}

The BDMS model we study is a multi-type birth-death process with mutations and incomplete sampling as presented by MacPherson et al.~\cite{macPhersonLoucaMcLauglinJoyPennell2021}.
It consists of a collection of lineages that give birth, die, mutate, and are sampled over time.
The model begins with a single lineage at time $\tau=\tmax > 0$ measured as the (positive) distance in time before the present day, $\tau=0$.
The single lineage at time $\tau=\tmax$ is initialized with type $a$ drawn from a categorical distribution with parameter $\pi=(\pi_1,\ldots,\pi_d)$ over a finite collection of $d$ types indexed by $1,\ldots,d$.
Each lineage progresses independently of all other lineages, with events arriving according to Poisson point processes with potentially time-varying intensity.
These processes are as follows.
At time $\tau$, a lineage of type $a$ gives birth to a lineage of type $b$ at rate $\lambda_{a,b}(\tau)$,
dies at rate $\mu_a(\tau)$,
and is sampled at rate $\psi_a(\tau)$.
Upon sampling, a lineage dies with probability $r_a(\tau)$.
When a lineage gives birth, there are two daughter lineages, one with type $a$ and one with type $b$.
The parameter $\lambda_{a,a}(\tau)$ describes the rate at which births without mutation occur.
For $b \neq a$,
the parameter $\lambda_{a,b}(\tau)$ describes the rate of cladogenetic mutation.
Anagenetic mutations---ones which occur at non-birth events---from type $a$ to type $b$ occur at rate $\gamma_{a,b}(\tau)$,
which is non-zero for only $b \neq a$.

In addition to the above processes, whose events arrive according to time-dependent Poisson point processes,
we allow for concerted sampling events (CSEs) \cite{macPhersonLoucaMcLauglinJoyPennell2021}.
These are instances in time at which a fraction of the population gets simultaneously sampled.
These occur at fixed times $t_0 \allowbreak  = 0 \allowbreak < t_1 \allowbreak  < \cdots \allowbreak  < t_L \allowbreak  \leq \tmax$.
The $l^\text{th}$ CSE consists of sampling all extant lineages of type $a$ at time $t_l$ independently with probability $\rho_{a,l}$.
During a CSE, each sampled lineage dies with probability $q_{a,l}$.
The model is defined by the set of parameters $\Theta = (\pi, \lambda, \mu, \gamma, \psi, r, \rho, q, t, \tmax)$,
whose interpretations are summarized in Table \ref{tab:bdms-summary}.
The rate parameters $\lambda,\mu,\gamma,\psi$ and the death probabilities $r$ may depend on time $\tau$.

\begin{table}
\begin{center}
\begin{tabular}{p{0.25\linewidth}p{0.75\linewidth}} \toprule
 Process & Description \\ \midrule
\textbf{Initialization} & At time $\tmax$, the population begins with one lineage with type $a \sim \text{Cat}(\pi)$. \\\addlinespace[.1cm]
\textbf{Birth} & For all $a,b\in\{1,\ldots,d\}$ and $\tau\in[0,\tmax]$, lineages of type $a$ split into types $a$ and $b$ at rate $\lambda_{a,b}(\tau)$ at time $\tau$. \\\addlinespace[.1cm]
\textbf{Death} & For all $a\in\{1,\ldots,d\}$ and $\tau\in[0,\tmax]$, lineages of type $a$ die at rate $\mu_a(\tau)$ at time $\tau$. \\\addlinespace[.1cm]
\textbf{Anagenetic mutation} & For all $a,b\in\{1,\ldots,d\}$ and $\tau\in[0,\tmax]$, lineages of type $a$ mutate to type $b$ at rate $\gamma_{a,b}(\tau)$  at time $\tau$.\\\addlinespace[.1cm]
\textbf{Sampling} & For all $a\in\{1,\ldots,d\}$ and $\tau\in[0,\tmax]$, lineages of type $a$ are sampled at rate $\psi_a(\tau)$ at time $\tau$. If sampled, they die with probability $r_a(\tau)$ at time $\tau$. \\\addlinespace[.1cm]
\textbf{CSEs} & For all $l\in\{0,\dots,L\}$ and $a\in\{1,\ldots,d\}$, lineages of type $a$ at time $t_l$ are sampled with probability $\rho_{a,l}$. If sampled, they die with probability $q_{a,l}$. \\\addlinespace[.1cm]
\bottomrule
\end{tabular}
\end{center}
\caption{Summary of processes in the birth-death-mutation-sampling (BDMS) model.}
\label{tab:bdms-summary}
\end{table}

\begin{figure}[t]
\centering
\includegraphics[width=0.9\linewidth]{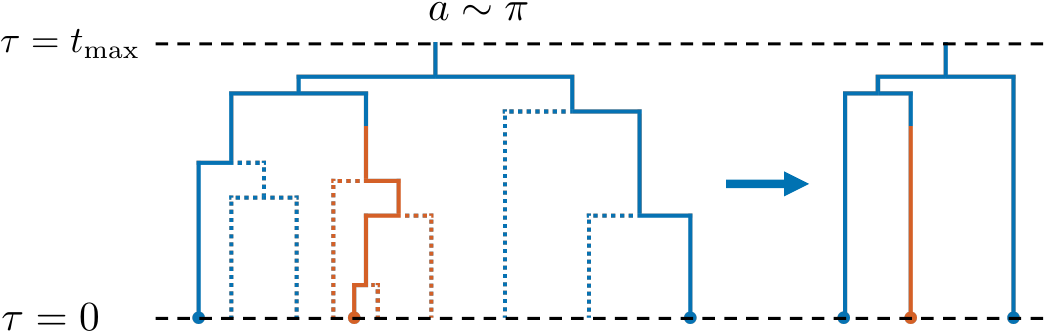}
\caption{Diagram of partial observation process. The full phylogeny is displayed on the left. Color represents \phenotype, with one anagenetic mutation event shown. All sampling events occur at the present day, represented by solid circles at the leaves. Lineages that are not ancestral to a sampled lineage are shown in dashed lines. The observed phylogeny is shown on the right.}
\label{fig:pruning-diagram}
\end{figure}

The reconstructed phylogeny refers to the subset of the full phylogeny containing only those lineages ancestral to a sampled lineage \cite{NeeMayHarvey1994}.
In Figure \ref{fig:pruning-diagram},
we provide a diagram demonstrating the partial observation process on a full phylogeny and the corresponding reconstructed phylogeny.
In applications, 
it is typically the reconstructed phylogeny (or an estimate of it based on sequence data) that is observed.
Thus, the distribution of the reconstructed phylogeny and its corresponding likelihood in terms of model parameters, rather than the distribution and likelihood for the full phylogeny, must be used as the basis for inference. 
A large body of work derives the likelihood for reconstructed phylogenies as a solution to a system of ordinary differential equations which can be solved numerically \cite{MaddisonMidfordOtto2007,MorlonParsonsPlotkin2011,LaudannoHaegemanRaboskyEtienne2020}.
Our simulation method will also rely on solving a system of ordinary differential equations.

\section{Simulation algorithms}

We describe two methods for simulating reconstructed phylogenies.
The first, which we call the \emph{full simulation}, simulates the full phylogeny and then removes (or ``prunes'') lineages that are not ancestral to a sampling event; this is the current state-of-the-art.
The second is our novel method, which we call the \emph{forward-equivalent simulation};  this approach simulates only those lineages that will have sampled descendants, thereby greatly reducing computational resources for most practical applications.

\subsection{The full simulation}
\label{sec:full-sim}

There are a variety of methods for simulating the full phylogeny which differ according to whether they grow the phylogeny in a depth-first or breadth-first manner and in how they determine the time and type of birth, death, mutation, and sampling events.
We used an efficient breadth-first implementation \cite{BDMS} whose time complexity is linear in the number of simulated events, where a single event corresponds to a birth, death, mutation, or sampling event in the simulated population.
Pseudocode for the algorithm we used is stated in Algorithm \ref{alg:full}.
The contribution of this paper is the forward equivalent simulation described in the next section, which can be viewed as a wrapper around the full simulation presented here.
Indeed, the forward equivalent simulation can be wrapped around any implementation of a time-varying BDMS model.
The implementation we used was selected to optimize time complexity in the number of simulated events.
While it is not the contribution of this paper, it is presented for completeness.
\begin{algorithm}[phbt!]
\caption{$\mathrm{FullSim}(\theta, \nmax)$}
\label{alg:full}
\begin{algorithmic}[1]
\State  $\tau \gets \tmax$; 
\State $S_a \gets \Call{RandomizedSet}$ for $a=1,\dots,d$;
 \State Draw $\mathrm{RootNode}$ with type $a \sim \text{Categorical}(\pi)$;
 \State $\Call{Insert}{\mathrm{RootNode},S_a}$;
 \While{$\tau \geq 0$ and $1 \leq \sum_{a=1}^d |S_a| \leq \nmax$}
  \State $\mathrm{EventType}$, $\tau_{\mathrm{event}}$ $\gets \Call{GetNextEvent}{\{|S_a|\},\tau}$
  \If{$\tau_{\mathrm{event}} < 0$}
  \ForAll{$a=1,\dots,d$ and $\mathrm{Node} \in S_a$}
  	\State $\mathrm{Child}\gets \Call{NewNode}{\mathrm{\phenotype}=a,\mathrm{event} = \mathrm{survival}, \mathrm{time} = 0} $;
  	\State $\Call{SetChildren}{\mathrm{Node},[\mathrm{Child}]}$;
  	\State $\Call{Remove}{\mathrm{Node},S_a}$; $\Call{Insert}{\mathrm{Child},S_a}$;\label{line:remove-insert}
  \EndFor
  \State{\text{Exit while loop}} 
  \EndIf
  \If{$\mathrm{EventType} = \mathrm{Birth}(a \rightarrow a,b)$}
   \State $\mathrm{Child}_1 \gets \Call{NewNode}{\mathrm{\phenotype}=a,\mathrm{event} = \mathrm{birth}, \mathrm{time} = \tau_{\mathrm{event}}}$; 
   \State $\mathrm{Child}_2 \gets \Call{NewNode}{\mathrm{\phenotype}=b,\mathrm{event} = \mathrm{birth},\mathrm{time} = \tau_{\mathrm{event}}}$;
   \State $\mathrm{Parent} \gets \Call{GetRandom}{S_a}$; \label{line:get-random}
   \State $\Call{SetChildren}{\mathrm{Parent},[\mathrm{Child}_1,\mathrm{Child}_2]}$;
   \State $\Call{Remove}{\mathrm{Parent},S_a}$;
   \State $\Call{Insert}{\mathrm{Child}_1,S_a}$; $\Call{Insert}{\mathrm{Child}_2,S_b}$;
  \EndIf
  \If{$\mathrm{EventType} = \mathrm{Death}(a)$}
   \State $\mathrm{Child} \gets \Call{NewNode}{\mathrm{\phenotype}=a,\mathrm{event} = \mathrm{death}, \mathrm{time} = \tau_{\mathrm{event}}}$;
   \State $\mathrm{Parent} \gets \Call{GetRandom}{S_a}$;
   \State $\Call{SetChildren}{\mathrm{Parent},[\mathrm{Child}]}$;
   \State $\Call{Remove}{\mathrm{Parent},S_a}$;
  \EndIf
  \If{$\mathrm{EventType} = \mathrm{Mutation}(a \rightarrow b)$}
   \State $\mathrm{Child} \gets \Call{NewNode}{\mathrm{\phenotype}=b,\mathrm{event} = \mathrm{mutation}, \mathrm{time} = \tau_{\mathrm{event}}}$;
   \State $\mathrm{Parent} \gets \Call{GetRandom}{S_a}$;
   \State $\Call{SetChildren}{\mathrm{Parent},[\mathrm{Child}]}$;
   \State $\Call{Remove}{\mathrm{Parent},S_a}$; $\Call{Insert}{\mathrm{Child},S_b}$;
  \EndIf
  \If{$\mathrm{EventType} = \mathrm{Sampling}(a)$}
   \State $\mathrm{Child} \gets \Call{NewNode}{\mathrm{\phenotype}=a,\mathrm{event} = \mathrm{sampling},\mathrm{time} = \tau_{\mathrm{event}}}$;
   \State $\mathrm{Parent} \gets \Call{GetRandom}{S_a}$;
   \State $\Call{SetChildren}{\mathrm{Parent},[\mathrm{Child}]}$;
   \State $\Call{Remove}{\mathrm{Parent},S_a}$; $\Call{Insert}{\mathrm{Child},S_a}$;
  \EndIf
  \State $\tau \gets \tau_{\mathrm{event}}$;
 \EndWhile
 \If{Capacity limit not exceeded}
 \ForAll{$a=1,\dots,d$ and $\mathrm{Node} \in S_a$}
  \If{$\mathrm{Bernoulli}(\rho_{a,0}) = 1$}
	  \State $\Call{SetEventType}{\mathrm{Node},\mathrm{event} = \mathrm{sampling}}$;
  \EndIf
 \EndFor
 \State Prune tree
 \EndIf\\
 \Return Tree $\cT$, or ``Empty'' if no nodes remain after pruning, or ``Capacity exceeded'' if capacity exceeded.
\end{algorithmic}
\end{algorithm}

In words, Algorithm \ref{alg:full} simulates forward in time starting at $\tau = \tmax$ and ending at $\tau = 0$ or until all lineages in the simulation die or a user-specified capacity limit $\nmax$ is exceeded.
The nodes in the tree correspond to birth, death, mutation, or sampling events.
As we grow the tree, each node contains information about its \phenotype, event type, and time.

We generate the tree in a breadth-first manner.
At each moment, the algorithm maintains biological time $\tau$ and the tree generated up until time $\tau$.
Each iteration generates the next event in biological time anywhere in the tree,
and advances the tree to the time of that next event.
Thus, consecutive iterations generate events that are consecutive in biological time but which need not occur to the same lineage or be close phylogenetically.

In order to achieve this, 
at each iteration the algorithm maintains, for each \phenotype $a = 1,\dots,d$,
a set $S_a$ containing the most recent event nodes for all extant lineages of \phenotype $a$.
The next event is then generated as follows:
\begin{enumerate}

    \item First, the time and type of the next event are determined. The type of the next event specifies i) whether it is a birth, death, mutation, or sampling event, ii)
    the \phenotype of the parent node, and iii)  for birth and mutation events, the \phenotype of the child or children nodes. It does not specify on which lineage the event occurs.
    In Algorithm \ref{alg:full},
    this step is represented by the function G{\footnotesize ET}N{\footnotesize EXT}E{\footnotesize VENT}.
    The time and type of the next event can be determined using the model parameters $\Theta$,
    the sizes of the extant populations for each \phenotype $\{|S_a|\}_{a=1}^d$,
    and the current biological time $\tau$.
    In \supptext{sec:further-implementation},
    we describe an implementation of G{\footnotesize ET}N{\footnotesize EXT}E{\footnotesize VENT} which runs in constant time in the size of the extant population.

    \item The parent node of the next event node is determined. If the event type returned by G{\footnotesize ET}N{\footnotesize EXT}E{\footnotesize VENT} specifies that the parent is a \phenotype $a$,
    we pick a parent node uniformly at random from $S_a$.
    As we describe below, we use a data structure for $S_a$ that permits uniform random sampling in $O(1)$ time.

    \item The phylogeny is advanced to the time of the next event. This involves creating a new event node whose parent is as chosen in the previous step and updating the sets $\{S_a\}$ as needed. See Algorithm \ref{alg:full} for details (which depend on the event type). This step requires possibly inserting or removing nodes from the sets $S_a$.
    As we describe below, we use a data structure for $S_a$ that permits insertion and removal in constant time.

\end{enumerate}
These iterations progress until either the time of the next event is after present time $\tau = 0$ or a user-specified population size capacity limit is exceeded.
If the time of the next event is after the present time,
then this event does not occur.
Instead, all extant lineages are advanced to the present,
and sampling occurs for each lineage according to the sampling probabilities $\rho_{a,0}$.

Note that we allow for a user-specified capacity limit even though this is not formally a part of the BDMS model.
Practically, it is useful to have an implementation that allows termination based on a user-specified capacity limit.
Depending on the model parameters, the population size can grow exponentially with time.
Thus, if $\tmax$ is set too large, the algorithm will fail to terminate in a reasonable amount of time.
In practice, one chooses model parameters such that simulation to the present is, except in rare cases, feasible with available computational resources.
The capacity limit is chosen large enough so that it rarely is reached but serves as protection against exponential blow-up.
In this case, the capacity limit will have a negligible impact on the distribution of simulated trees.

After the full phylogeny has been simulated as described above, the tree is pruned to remove all lineages and nodes that are not ancestral to a sampling event.
Moreover, birth events in which only one child lineage survives either (1) are removed if the surviving offspring has the same phenotype as the parent, or (2) appears as anagenetic mutations if the surviving offspring has a different phenotype than the parent.
The pruning operation can be carried out in time linear in the number of events in the full phylogeny.

For simplicity and conciseness,
we have omitted from this description and from the pseudo-code in Algorithm \ref{alg:full}  concerted sampling events (CSEs) except for the one at the present day.
These are straightforward to add and do not affect our discussion on computational complexity.
Indeed, when the time of the next event is determined to be after the next concerted sampling event, the event does not occur. Instead, all extant lineages are advanced to the time of the concerted sampling event, at which point the appropriate sampling and death events are carried out.

We implement Algorithm \ref{alg:full} so that each iteration of the while loop (which corresponds to a single event) runs in constant time in the population size.
This requires that G{\footnotesize ET}N{\footnotesize EXT}E{\footnotesize VENT} has constant time complexity $O(1)$ in the population size, and  the sets $S_a$ are implemented using a data structure which supports insertion, removal, and uniform random sampling in constant time in the population size.
We describe these implementation details in \supptext{sec:further-implementation}.
With this implementation,
Algorithm \ref{alg:full} has time complexity linear in the number of events in the full phylogeny.
In Section \ref{sec:empirical} and \suppfref{fig:time-vs-unpruned-size},
we empirically validate that the computation time scales linearly with the number of events in the full phylogeny.

As we describe in \supptext{sec:further-implementation},
the time complexity of G{\footnotesize ET}N{\footnotesize EXT}E{\footnotesize VENT} is quadratic in the size of the \phenotype space $d$.
Thus, the overall runtime of Algorithm \ref{alg:full} when taking into account the size of the \phenotype space is $O(\#\text{nodes}(\cT)d^2)$ where $\#\text{nodes}(\cT)$ is the number of events in the full phylogeny $\cT$.
The quadratic factor $d^2$ may be replaced by $d + \log(\#\text{nodes}(\cT))$ by using an alternative implementation based on a priority queue.
For applications with large \phenotype spaces, this implementation may be preferred.
We describe this approach in \supptext{sec:priority-queue}.
In this work, however, we focus on the approach provided in Algorithm \ref{alg:full}, since its runtime is strictly linear in the number of events in the full phylogeny.

\subsection{The forward-equivalent simulation}
\label{sec:FE-sim}

The full simulation expends computational time on all lineages, including those that have no sampled descendants and thus do not appear in the reconstructed phylogeny.
We propose a method that does not expend computational resources on such unobserved lineages.
Avoiding such computation is non-trivial:
the appearance of a lineage in the reconstructed phylogeny depends on the occurrence (or not) of a future sampling event.
Thus, in Algorithm \ref{alg:full},
the appearance of an event in the reconstructed phylogeny cannot be determined at the time the event is simulated.

In this section,
we show that the distribution of the reconstructed phylogeny in the full model conditional on it being non-empty is equal to that generated by an alternative BDMS model without death and with complete sampling at the present.
We call this alternative BDMS model the \emph{forward-equivalent model}.

Precisely, consider any set of model parameters $\Theta := (\pi, \lambda, \mu, \gamma, \psi, r, \rho, q, t, \tmax)$ for the full simulation.
Our main result is that, for any such $\Theta$, there is an alternative set of parameters $\Theta^\FE := (\pi^\FE,\allowbreak\lambda^\FE,\allowbreak\mu^\FE,\allowbreak\gamma^\FE,\allowbreak\psi^\FE,\allowbreak r^\FE,\allowbreak\rho^\FE,\allowbreak t,\allowbreak\tmax)$ satisfying $\mu_a^\FE(\tau) = 0$ for all $a\in \{1,\dots,d\}$, $\tau \in [0,\tmax]$ (no death) and $\rho_{a,0}^\FE = 1$ for all $a\in \{1,\dots,d\}$ (complete sampling at the present) such that the distribution of the reconstructed phylogeny $\cT$ under $\Theta$ conditional on being non-empty, $\cT \ne T_\emptyset$, is equal to the distribution of the reconstructed phylogeny under $\Theta^\FE$ (with $T_\emptyset$ denoting the empty tree realization, which corresponds to a full tree in which all lineages are either extinct or unsampled, so that pruning leaves no ascertained subtree).
The parameters $\Theta^\FE$ are given precisely in Section \ref{sec:FE-model} via a map $F$ on parameter space such that $\Theta^\FE = F(\Theta)$.
Let random variable $\cT_\Theta$ denote the reconstructed phylogeny in the full model parameterized by $\Theta$.
\begin{theorem}
\label{thm:forward-equivalence}
	The distribution of the reconstructed phylogeny in the full model conditional on its non-emptiness is equivalent to the unconditional distribution of the reconstructed/full phylogeny in the forward-equivalent model:
	\begin{equation}
	\label{eq:forward-equivalence}
		\Law(\cT_\Theta \mid \cT_\Theta \ne T_\emptyset) = \Law(\cT_{F(\Theta)}).
	\end{equation}
\end{theorem}
\noindent 
We prove Theorem \ref{thm:forward-equivalence} in \supptext{sec:FE-proof}.

Theorem \ref{thm:forward-equivalence} suggests simulating the reconstructed phylogeny in the full model by running the full simulation on the parameters $\Theta^\FE = F(\Theta)$.
This approach, which we call the forward-equivalent simulation,
is given in Algorithm \ref{alg:FE}.
It generates reconstructed phylogenies from the distribution $\Law(\cT_\Theta \mid \cT_\Theta \ne T_\emptyset)$.
If one wants to generate reconstructed phylogenies from $\Law(\cT_\Theta)$ allowing for emptiness,
one can first generate a Bernoulli random variable $B$ with success probability $\P(\cT_\Theta \neq T_\emptyset)$ (which is computed in Section \ref{sec:FE-model}) and,
if $B = 0$, return $T_\emptyset$, 
whereas if $B = 1$, return $\mathrm{FESim}(\Theta,\nmax)$.
In practice, one often will want to generate non-empty phylogenies, so this extra step will not be necessary.

\begin{algorithm}[hbt!]
\caption{$\mathrm{FESim}(\Theta,\nmax)$}
\label{alg:FE}
\begin{algorithmic}[1]
\State $\Theta^\FE \gets F(\Theta)$\label{line:evaluate-F} \Comment{Precomputated and reused for sampling multiple trees}
\State $\cT \gets \text{FullSim}(\Theta^\FE,\nmax)$\label{line:call-full-sim}
\State \Return $\cT$
\end{algorithmic}
\end{algorithm}

In the forward-equivalent model,
all lineages are observed and the full and the reconstructed phylogeny are the same.
When used to simulate reconstructed phylogenies in models with death or incomplete sampling,
the forward-equivalent simulation effectively avoids expending computational resources on lineages that are not observed.
This leads to substantial computational savings when either only a small fraction of the full phylogeny is observed or the full phylogeny only rarely survives until the present.
An analysis of computational complexity is carried out in Section \ref{sec:complexity}.

Implementing Algorithm \ref{alg:FE} requires evaluating the forward-equivalent model parameters $\Theta^\FE$ via the map $F$ on line \ref{line:evaluate-F}.
These can be pre-computed once,
and then used to generate an arbitrary number of trees via repeated calls to the full simulation, as on line \ref{line:call-full-sim}.
These precomputations are described in \supptext{sec:further-implementation}.
The cost of precomputation is amortized over the number of simulated trees,
so contributes negligible computational cost when this number is large.
In our experiments, pre-computations were very fast.

In practice, one may choose to set $n_{\text{max}} < \infty$ in Algorithm \ref{alg:FE}.
Because in the forward-equivalent model, reconstructed and full phylogenies are the same, simulating from the forward equivalent model with finite $n_{\text{max}}$ is equivalent to simulating from the full model conditional on its non-emptiness and on the maximal number of simultaneously extant lineages at any point in time in the reconstructed phylogeny being bounded above by $n_{\text{max}}$.

\subsection{The forward-equivalent model parameters}
\label{sec:FE-model}

In this section,
we define the map $F$ giving parameters $\Theta^\FE$ explicitly in terms of the parameters $\Theta$ of the full model.
A key quantity is the non-observation probability $E_a(\tau)$ for $a=1,\dots,d$ and $\tau \in [0,\tmax]$ in the full model.
It is the probability that, given a lineage of type $a$ in the population at time $\tau$, none of its descendants will be sampled. 
 Note that
$\P(\cT_\Theta \neq T_\emptyset)$ is given by $1-\sum_{a=1}^d \pi_a E_a(\tmax)$.
At CSEs, the non-observation probability will be different immediately prior to and after the realization of the CSE's sampling and death events.
We denote the non-observation probability prior to the realization of these events by $E_a(t_l^+)$ and after the realization of these events by $E_a(t_l^-)$, where $t_l$ is the time of the CSE.
As shown in \cite[Equation~15]{macPhersonLoucaMcLauglinJoyPennell2021}),
the non-observation probabilities are given by the solution to the following ODE system for all $a=1,\dots,d$ and $\tau\in[0,\tmax]$:
\begin{equation}
\label{eq:reconstructed-E-ODE}
\begin{aligned}
	&\textbf{Initialization:}&E_a(0^-) = &1,
	\\
	&\textbf{Between CSEs:}&\frac{\de E_a(\tau)}{\de \tau}
		=
		&\sum_{b=1}^d\lambda_{a,b}(\tau)[E_a(\tau)E_b(\tau) - E_a(\tau)]
	\\
		&&&+
		\mu_a(\tau)[1 - E_a(\tau)]
	\\
		&&&+
		\psi_a(\tau)[0 - E_a(\tau)]
	\\
		&&&+
		\sum_{b=1}^d \gamma_{a,b}(\tau)[E_b(\tau) - E_a(\tau)],
	\\
	&\textbf{At CSEs:}&E_a(t_l^-) &= \lim_{\tau \uparrow t_l} E_a(\tau),
		\quad E_a(t_l^+) = (1 - \rho_{a,l})E_a(t_l^-).
\end{aligned}
\end{equation}
For completeness, we derive \eqref{eq:reconstructed-E-ODE} in \supptext{sec:FE-proof}.
The non-observation probability has previously played an important role in likelihood-based and Bayesian methods of inference based on the reconstructed phylogeny \cite{MaddisonMidfordOtto2007,FitzJohn2009,Hohna2011,Morlon2014,neherRusselShraiman2014,macPhersonLoucaMcLauglinJoyPennell2021}.
For all $a,b\in\{1,\dots,d\}$ and $\tau\in[0,\tmax]$,
the forward-equivalent mapping $\Theta^\FE = F(\Theta)$ is defined by 
\begin{equation}
\label{eq:FE-params}
\begin{aligned}
	\pi_a^\FE
		&=
		\frac{\pi_a[1-E_a(\tmax)]}{\sum_b \pi_b[1-E_b(\tmax)]},
	&
	\lambda_{a,b}^\FE(\tau)
		&=
		[1 - E_b(\tau)]\lambda_{a,b}(\tau),
	\\
	\mu_a^\FE(\tau)
		&=
		0,
	&
	\psi_a^\FE(\tau)
		&=
		\frac{\psi_a(\tau)}{1 - E_a(\tau)},
	\\
	r_a^\FE(\tau)
		&=
		r_a(\tau) + [1-r_a(\tau)]E_a(\tau),
	&
	\gamma_{a,b}^\FE(\tau)
		&=
		\frac{1 - E_b(\tau)}{1 - E_a(\tau)}
		\Big[
			\gamma_{a,b}(\tau)
			+
			E_a(\tau)\lambda_{a,b}(\tau) \delta_{a \neq b}
		\Big],
	\\
	\rho_{a,l}^\FE
		&=
		\frac{\rho_{a,l}}{1 - E_a(t_l^+)},
	&
	q_{a,l}^\FE
		&=
		q_{a,l} + (1-q_{a,l})E_a(t_l^-).
\end{aligned}
\end{equation}
By \eqref{eq:reconstructed-E-ODE}, $E_a(0^+) = 1 - \rho_{a,0}$,
whence by \eqref{eq:FE-params}, $\rho_{a,0}^\FE = 1$.
Thus, in the forward-equivalent simulation,
all lineages at $\tau = 0$ are sampled.
We note that the ODE \eqref{eq:reconstructed-E-ODE} is Lipshitz continuous in its solution on each inter-CSE interval, so for $\lambda,\mu,\gamma,\psi,r$ suitably smooth in the time domain, it follows by standard results that there exists a unique solution, i.e., the map $F$ is well-defined.
It suffices, for example, to assume that the parameters are continuous or piecewise constant in time.

It is enlightening to interpret some of the equations in \eqref{eq:FE-params}.
The equation for $\lambda_{a,b}^\FE(\tau)$ indicates that the birth rate in the reconstructed phylogeny is smaller than that in the population.
This effect is stronger for individuals that are more unlikely to have sampled descendants.
The equation for $\gamma_{a,b}^\FE(\tau)$ indicates that anagenetic mutation rates can be upward or downward biased in the reconstructed phylogeny. 
In particular, mutations from \phenotypes that are unlikely to have sampled descendants and to \phenotypes that are likely to have sampled descendants are higher in the reconstructed phylogeny than in the population; mathematically, $E_a(\tau) > E_b(\tau)$ implies $\frac{1-E_b(\tau)}{1-E_a(\tau)}>1$, so $\gamma_{a,b}^\FE(\tau) > \gamma_{a,b}(\tau)$.
One factor that increases the probability of having a sampled descendant is having ``higher fitness'' in terms of a larger birth rate or lower death rate, leading to a higher chance of survival.
Thus, these equations indicate that the relative birth rates of ``fit'' vs.\ ``unfit'' individuals in the reconstructed phylogeny are higher than in the population,
and mutation rates from ``unfit'' to ``fit'' individuals are enhanced.
These biases exemplify the importance of incorporating the sampling process into phylogenetic simulations.

\begin{figure}[H]
\centering
\includegraphics[width=0.5\textwidth]{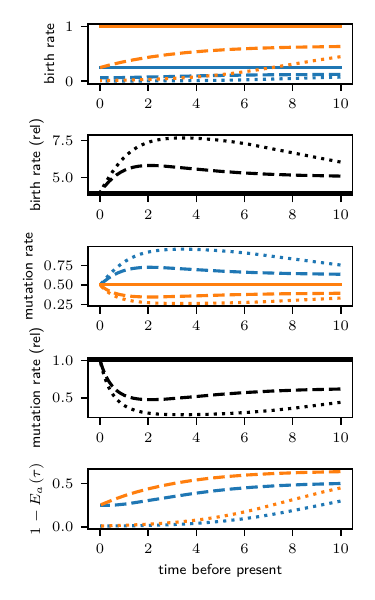}\\
\includegraphics[width=0.5\textwidth]{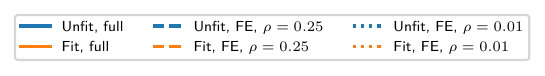}
\caption{Full and forward equivalent model parameters as a function of time in a two-type model with a $\Fit$ and $\Unfit$ phenotype,
with $\lambda_\Fit = 1.0$, $\lambda_\Unfit = 0.25$, $\mu_\Fit = \mu_\Unfit = 0.25$, and anagenetic mutation rates $\gamma_{\Fit,\Unfit} = \gamma_{\Unfit,\Fit} = 0.5$.
The relative birth rate is defined as $\lambda_\Fit(\tau)/\lambda_\Unfit(\tau)$ and $\lambda_\Fit^\FE(\tau)/\lambda_\Unfit^\FE(\tau)$.
The relative mutation rate is is defined as $\gamma_{\Fit,\Unfit}(\tau)/\gamma_{\Unfit,\Fit}(\tau)$ and $\gamma_{\Fit,\Unfit}^\FE(\tau)/\gamma_{\Unfit,\Fit}^\FE(\tau)$.}
\label{fig:FE-parameters}
\end{figure}

As a concrete example, we plot in Figure \ref{fig:FE-parameters} the forward equivalent model parameters in a two-type model with a $\Fit$ and $\Unfit$ phenotype,
with $\lambda_\Fit = 1.0$, $\lambda_\Unfit = 0.25$, $\mu_\Fit = \mu_\Unfit = 0.25$, and anagenetic mutation rates $\gamma_{\Fit,\Unfit} = \gamma_{\Unfit,\Fit} = 0.5$.
We consider both a low sampling probability $\rho = 0.01$ and moderate sampling probability $\rho = 0.25$, occuring uniformly across phenotypes at the present.
We observe that the forward equivalent birth rate is smaller than that in the population for both the Fit and Unfit phenotypes.
Moreover, we see that in the forward equivalent model, birth rates are higher in the distant past than in the present.
This heightened diversification rate in the observed phylogeny in the distant past has been termed the ``push of the past'' \cite{Budd2018-on} and results from survivorship bias.
We also plot the relative birth rate in the forward equivalent model, defined as $\lambda_\Fit^\FE(\tau) / \lambda_\Unfit^\FE(\tau)$. We see that it is larger than 4, indicating that the birth rate of the Unfit phenotype is reduced more relative to its population value than that of the Fit phenotype.
This effect is stronger when $\rho$ is smaller.
These qualitative phenomena can be easily seen in the formulas in \eqref{eq:FE-params}.
Indeed, the amount by which the birth rate is depressed is controlled by $1 - E_b(\tau)$, the probability a lineage will end up in the sample,
so that the qualitative time dependence of the birth rates,
the inflation of the relative birth rate,
and the dependence of the relative birth rate on $\rho$ can be read off the plot of $1 - E_a(\tau)$ in Figure \ref{fig:FE-parameters}.
We also observe that the mutation rate from the Unfit to Fit phenotype is higher and from the Fit to Unfit phenotype is lower in the forward equivalent model than in the population,
with a stronger effect for smaller $\rho$.
This is the result of selective pressures, which are more pronounced when $\rho$ is smaller.
Again, \eqref{eq:FE-params} indicates the relevant quantity is the ratio $(1-E_b(\tau))/(1-E_a(\tau))$,
the qualitative behavior of which can also be observed in the plot of $1 - E_a(\tau)$ in Figure \ref{fig:FE-parameters}.

\subsection{Efficiency improvements of forward-equivalent simulations}
\label{sec:complexity}

As we justified in Section \ref{sec:full-sim},
our simulation algorithm has time complexity which is linear in the number of simulated events.
The forward-equivalent simulation can be faster than the full simulation because it simulates fewer events per non-empty reconstructed phylogeny.

There are two reasons why the forward-equivalent simulation simulates fewer events per non-empty reconstructed phylogeny.
First, the reconstructed phylogeny contains fewer events (i.e., nodes) than the full phylogeny.
For example, in a pure birth process that samples present-day lineages with probability $\rho$, 
the reconstructed tree will have approximately $\rho$ times as many nodes as the full phylogeny. We thus expect the forward-equivalent simulation to run $1/\rho$ times faster per tree than the full simulation.

Second, in the full simulation, the reconstructed phylogeny may be empty.
Thus, the full simulation may require multiple retries to produce a single non-empty reconstructed phylogeny.
The events simulated in trees with empty reconstructed phylogenies contribute to the computational cost of simulating a single non-empty phylogeny.
In contrast, the forward-equivalent simulation never produces the empty tree $T_\emptyset$.
As an artificial extreme case,
consider a trivial process without birth or death and which samples present-day lineage with probability $\rho$.
In this case, the full tree is a `stick'.
The reconstructed tree is a `stick' with probability $\rho$ and is $T_\emptyset$ otherwise.
The forward equivalent simulation simulates a stick in its first try, requiring $\Theta(1)$ time per non-empty reconstructed tree.
The full simulation requires in expectation $1 / \rho$ retries to obtain a non-empty reconstructed tree.
In this case, the speedup provided by our method is again $1/\rho$.

In what follows, we study the key quantity $R$ (respectively, $R^\FE$),
which is the expected number of simulated events required to produce a single non-empty reconstructed tree in the full (respectively, forward-equivalent) simulation.
For a single-type birth-death process, we prove the following result:
\begin{proposition}
\label{proposition:bd-speedup}
	Consider running a birth-death process for 1 unit of time with birth rate $\lambda$, death rate $\mu$, and extant sampling probability $\rho$.
	Then:
    \begin{enumerate}
        \item If $\lambda = \mu$, then the theoretical speedup of the forward-equivalent model is:
        \begin{equation*}
            \frac{R}{R^\FE} = \frac{1}{\rho}(1 + \mu).
        \end{equation*}
        \item If $\lambda \neq \mu$, then the theoretical speedup of the forward-equivalent model is:
        \begin{equation*}
            \frac{R}{R^\FE} = \frac{1}{\rho}\Bigg[1 + \frac{\mu(e^{\lambda - \mu} - 1)}{(\lambda - \mu)e^{\lambda - \mu}}\Bigg].
        \end{equation*}
    \end{enumerate}
\end{proposition}
\noindent 
We prove Proposition \ref{proposition:bd-speedup} in \supptext{sec:speedup}.
In the simpler case that the birth and death rates are equal, we see that the theoretical speedup is essentially linear in the (inverse) sampling probability $\rho$ and in the death rate $\mu$.
Later, in our empirical benchmarks, we will see that this theoretical estimate is quite accurate.

For the general BDMS model, we have the following result.
\begin{proposition}
\label{prop:general-speedup}
    Let $\cT$ denote the full tree in the full model,
    and $\cR(\cT)$ its corresponding (possibly empty) reconstructed tree.
    Then
    \begin{equation*}
        \frac{R}{R^\FE} = \frac{\E[\#\text{nodes}(\cT)]}{\E[\#\text{nodes}(\cR(\cT))]}.
    \end{equation*}
\end{proposition}
\noindent 
We emphasize here that nodes in $\cT$ include birth events, mutation events, and sampling events and thus need not all be locations of bifurcation in the tree.
We prove Proposition \ref{prop:general-speedup} in \supptext{sec:speedup}.
The significance of Proposition \ref{prop:general-speedup} is that it unifies the two sources of speed-up described above,
avoiding explicit accounting of the number of retries required to observe a non-empty reconstructed phylogeny.

We remark that the expected number of nodes appearing in Proposition \ref{prop:general-speedup} can be computed numerically with time discretization and dynamic programming.
Thus, Proposition \ref{prop:general-speedup} allows for numerical estimates of speed-up factors.
Empirically,
the forward-equivalent simulation may experience a constant factor computational overhead relative to the full simulation due to its sampling from time-inhomogeneous Poisson processes: see Section \ref{sec:efficiency-experiments}.
In our experiments, this overhead was smaller than a factor of 2, so the forward-equivalent simulation was faster than the full simulation for moderate values of $\rho$ and $\mu$.

\section{Empirical validation}
\label{sec:empirical}

We carry out several simulations to validate our method.
In Section \ref{sec:dist-test},
we consider a case where both the full and forward-equivalent simulations are feasible,
and perform several tests of whether the reconstructed phylogenies produced by the two simulations have the same distribution.
In Section \ref{sec:efficiency-experiments},
we perform several simulations to measure the relative computational efficiency of the two approaches and to validate the theory from Section \ref{sec:complexity}.
In Section \ref{sec:massive-sim},
we use the forward-equivalent simulation to generate phylogenies from a setting in which the full simulation is computationally infeasible on our machines.

\subsection{Distribution tests}
\label{sec:dist-test}

We performed a simulation study to test that the full and forward-equivalent simulations produce reconstructed phylogenies with the same distribution. 
We consider a full model consisting of two \phenotypes, $\Fit$ and $\Unfit$. 
The population is simulated for $\tmax = 20$ units of biological time. 
Both \phenotypes have a death rate of $\mu_\Fit = \mu_\Unfit = 0.25$, the $\Unfit$ \phenotype  has birth rate of $\lambda_\Unfit = 0.25$, and the $\Fit$ \phenotype  has a birth rate of $\lambda_\Fit = 1$.
Recall that, without mutation and given infinite time, a population will go extinct with probability 1 if and only if the birth rate does not exceed the death rate: $\lambda \leq \mu$.
Thus, we have chosen parameters so that the $\Unfit$ population is just barely in the regime where it would, without mutation, be assured eventual extinction.
On the other hand, the $\Fit$ population would, without mutation, be unlikely to go extinct and would grow exponentially fast.
We consider anagenetic mutation rates $\gamma_{\Fit,\Unfit} = 0.8$ and $\gamma_{\Unfit,\Fit} = 0.1$.
Thus, deleterious mutations from $\Fit$ to $\Unfit$ are much more common than beneficial mutations from $\Unfit$ to $\Fit$. 
Based on our simulations (see below),
these mutation rates keep the distribution of \phenotypes in the population relatively balanced.
We consider a model with sampling at the present $\rho_{\Fit,0} = \rho_{\Unfit,0} = 0.5$ and no sampling at earlier times or at death events.

We generated 1,000 non-empty phylogenies from both the full and forward-equivalent simulations for the above model.
Because we are conditioning on non-emptiness,
generating a single non-empty phylogeny in the full simulation may (and sometimes did) require multiple retries.

We tested distributional equivalence of several scalar summary tree statistics.
Because phylogenetic trees are high-dimensional objects,
distributional agreement on these test statistics does not necessarily imply distributional equivalence of the phylogenetic trees.
Moreover, the tree statistics we selected are not intended to be exhaustive,
and we do not intend to imply that they provide an optimal summary of tree geometry for any particular application.
Rather, we intend to capture a wide enough range of tree features to provide convincing evidence that the distributional equivalence \eqref{eq:forward-equivalence} holds.
The tree statistics we considered were:
\begin{itemize}

	\item \textbf{Event count.} The total number of nodes in the tree, which includes births, mutations, and sampled leaf nodes.

	\item \textbf{Leaf count.} The total number of leaves in the tree. In this simulation, this is the size of the sample from the present-day population.

	\item \textbf{Branch length.} The cumulative branch length stratified by \phenotype (for $\Fit$ and $\Unfit$ lineages) and in total.

	\item \textbf{Subtree count, stratified by size.} The number of nodes in the tree for which the descendant tree has \textsf{size} nodes (including the root node). For example, the subtree count at $\textsf{size} = 1$ is the number of leaves. The subtree count at $\textsf{size} = 3$ contains the sum of the number of cherries and the number of nodes after which one mutation event but no birth events occur before present-day sampling.

	\item \textbf{Lineage count, stratified by \phenotype and time.} For a \phenotype in $\{\Fit,\Unfit\}$ and time $t$, the number of lineages in the tree of that \phenotype at a cross-section at time $t$.

\end{itemize}

\noindent We also considered the distribution of ``block statistics''.
In multi-type models, each phylogeny is partitioned into a collection of maximally connected subtrees of uniform \phenotype.
We call each element of this partition a \textit{block}.
A block statistic is a scalar summary of a block; for example, total branch length.
We simulated 1,000 non-empty trees and then plot histograms of various block statistics over the list of generated blocks.
We emphasize that not all trees contain the same number of blocks,
so that trees with more blocks contribute more counts to these histograms than trees with fewer blocks.
The histograms represents a distribution which assigns an equal weight to each block, and thus assigns a weight to each tree proportional to its number of blocks. 
We consider the following block statistics:
\begin{itemize}

	\item \textbf{Events per block, stratified by \phenotype.} The number of nodes in the block.

	\item \textbf{Branch length per block, stratified by \phenotype.} The cumulative branch length of the block.

\end{itemize}

\begin{figure}[t!]
\centering
\includegraphics[width=\textwidth]{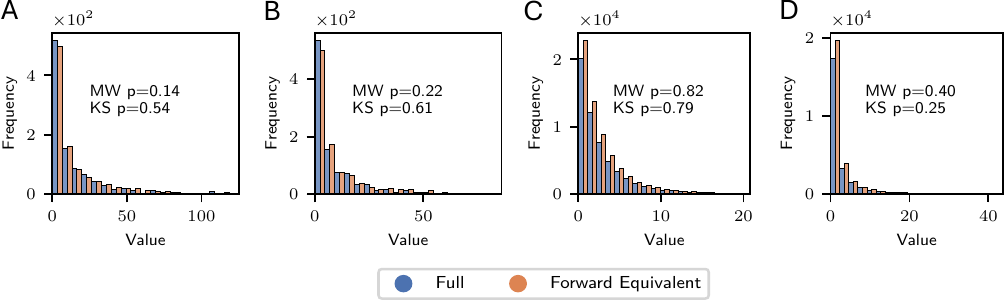}
\caption{Distributional comparison of summary statistics from two-type model described in the text. The full and forward equivalent simulation are both run until 1,000 non-empty trees are generated. Each plot displays a different summary statistics and the Mann-Whitney (MW) and Kolmogorov-Smirnov (KS) p-value testing the equivalence of the full and forward equivalent simulations. (A) Number of extant lineages at biological time $t = 16$ with phenotype $\Unfit$. (B) Number of extant lineages at biological time $t = 16$ with phenotype $\Fit$. (C) Branch length of blocks with phenotype $\Unfit$. (D) Branch length of blocks with phenotype $\Fit$.
The simulation parameters are described in the main text, Section \ref{sec:dist-test}.
}
\label{fig:dist-test}
\end{figure}

\noindent In Figure \ref{fig:dist-test},
we show histograms for two of these summary statistics---lineage count and branch length per block---stratified by \phenotype.
For the block-based statistics (in this case, branch length per block), the $y$-axis counts the number of blocks cumulatively across phylogenies, so that each phylogeny may contribute multiple blocks to the count, and larger phylogenies will typically contribute more.
In all cases, the histograms generated by the full and forward-equivalent simulations are visually similar.
For each summary statistic, we performed a two-sided Mann-Whitney U-test and the Kolmogorov-Smirnov test to compare the distributions, with the $p$-values in each case reported on the corresponding histogram plot.
We caution that the theoretical null distribution used to compute Kolmogov-Smirnov test statistic is only asymptotically valid for continuous distributions.
Thus, it must be interpreted with caution for the count-based test statistics when the counts are small.
Similar plots to Figure \ref{fig:dist-test},
reporting results for all the summary statistics listed above, are provided in \suppfref{fig:dist-comp}.

While not a proof of distributional equivalence, these tests indicate that our implementation successfully produces the distributional equivalence predicted by Theorem \ref{thm:forward-equivalence}, providing evidence for its correctness.

\subsection{Computational efficiency}
\label{sec:efficiency-experiments}

We empirically assessed the relative computational cost of the full and the forward-equivalent simulations and their dependence on population size and the size of the sub-sampled phylogeny under three models: single-type without death, single-type with death, and two-type.

The ``single-type without death'' model consisted of a unique \phenotype with constant birth rate given by $\lambda = 1.0$ and no death run for $\tmax = 5$ units of biological time.
We subsampled the present population with probability $\rho$, which we set to $ \rho = 0.1 \times k$ for $k  = 1,\ldots,10$.
At each value of $k$, we simulated 50 non-empty trees.
As in Section \ref{sec:dist-test}, the full simulation may require multiple retries to generate a single non-empty tree.

\begin{figure}[t]
\centering
\includegraphics[width=0.5\textwidth]{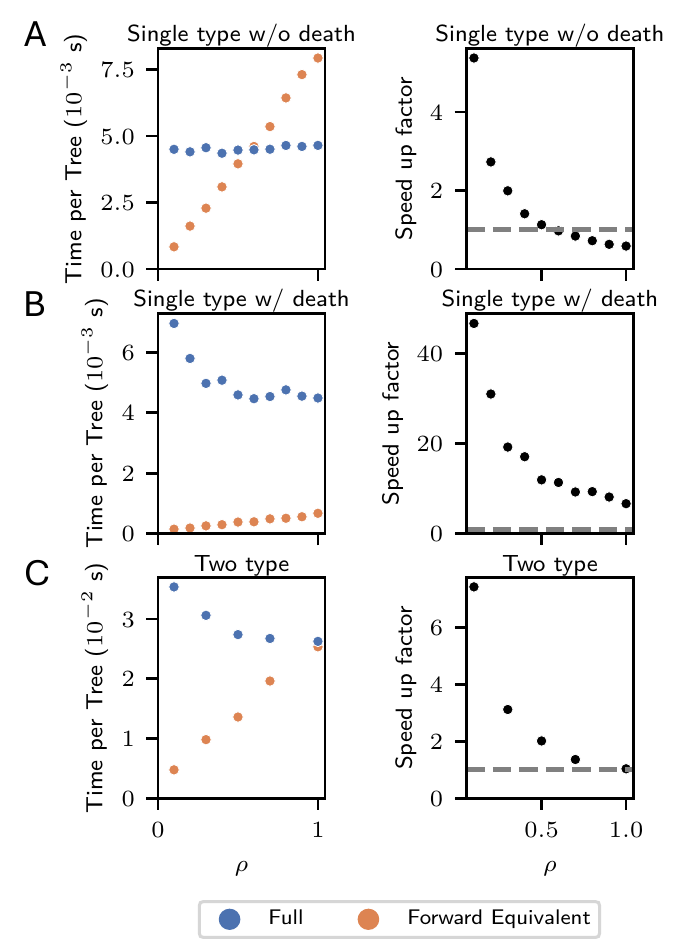}
\caption{Comparison of computational costs. Time per tree (s) is the total amount of simulation time required to produce 50 non-empty trees, divided by 50. Speed up factor is the ratio of the time per tree from the full simulation to the forward equivalent simulation. All sampling occurs at present with probability $\rho$ for each extant lineage identically across types. Horizontal dashed line at $\text{speed up factor} = 1$. A: single-type without death. $\lambda = 1.0$, $\mu = 0$, $\tmax = 5$. B: single-type with death. $\lambda = 1.0$, $\mu = 1.0$, $\tmax = 10$. C: two-type with death. Two types: Fit and Unfit. $\lambda_{\Fit} = 1.0$, $\lambda_\Unfit = 0.25$, $\mu_\Fit = \mu_\Unfit = 0.25$.}
\label{fig:timing-figure}
\end{figure}

All results are displayed in Figure \ref{fig:timing-figure}.
In the left column
we report, for each value of $\rho$, the average total simulation time (including retries) required to produce 50 non-empty trees, normalized by 50.
In the right column,
we report the ratio of the computation time required by the full simulation to the forward-equivalent simulation on a log-log scale.

The first row shows results for the single-type model without death. 
For the full simulation,
the mean time per tree is approximately constant in $\rho$.
For the forward-equivalent simulation,
the mean time per tree appears to be linear in $\rho$.
This is consistent with our simulation scaling linearly in the total number of simulated events,
which in the full simulation is independent of $\rho$ and in the forward-equivalent simulation is proportional to $\rho$ on average.

The speed-up factor is decreasing with $\rho$,
and is larger than $1$ for $\rho \leq 0.5$,
reaching $\approx 5$ at $\rho = 0.1$.
Regressing the log speed up factor on $\log \rho$ gives the approximation $\textsf{speed up factor} \approx 0.58  \rho^{-0.99}$,
consistent with the inverse dependence on $\rho$ predicted by Proposition \ref{proposition:bd-speedup}.
The speed-up factor is smaller by approximately a factor of 2 relative to the theory.
We believe this is because, in this model, the forward-equivalent incurs some overhead relative to the full simulation because it must sample from time-inhomogenous rather than constant-rate Poisson processes (see \supptext{sec:further-implementation} for implementation details).
For $\rho \gtrsim 0.6$, the gain from avoiding simulating unobserved events does not overcome this computational overhead,
and the speed-up factor is less than one.
We emphasize, however, that this computational overhead increases computation time by a constant factor which does not grow with phylogeny size.
Thus, it will be overcome by a small amount of subsampling even for very large phylogenies.
For models that do not include death, the forward-equivalent simulation is faster than the full simulation when there is at least a moderate amount of subsampling.
When the full model itself includes time-inhomogenous rate parameters,
the forward-equivalent simulation will not have this computational overhead relative to the full simulation.

In the second row of Figure \ref{fig:timing-figure}, we present similar results for a single-type model that includes death.
Precisely, we consider a regime with birth rate $\lambda = 1.0$ and death rate $\mu = 1.0$, run for $\tmax = 10$ units of biological time.
The computational savings are even more dramatic in this case, 
at $\rho = 0.1$ reaching a speed-up factor of $\approx 45$ 
and at $\rho = 1$ a speed-up factor of $\approx 6$.
In this case, even at $\rho = 1$, the forward-equivalent simulation is much faster than the full simulation.
This is because lineages that die are not sampled.
For models that include death, the forward-equivalent simulation can be more efficient for all sub-sampling schemes.

We also observe that the computation time for the full simulation is no longer approximately constant
in the sampling probability $\rho$. In fact, unlike in the forward-equivalent simulation, the computation tends to be larger for smaller values of $\rho$.
We suspect this is a result of the computational cost of pruning the phylogeny,
which is more computationally intensive when a smaller fraction of the tree is sampled.

Finally, in the third row of Figure \ref{fig:timing-figure},
we present similar results for the two-type model described in Section \ref{sec:dist-test},
and we observe similar phenomena to that appearing in the single-type simulations.

In \suppfref{fig:time-vs-unpruned-size},
we present further plots demonstrating the computational overhead described above and the linear scaling of computation time in the total number of simulated events.

\subsection{Subsampling from massive phylogenies}
\label{sec:massive-sim}

The previous section focused on situations in which, although the forward-equivalent simulation was much faster than the full simulation, the phylogeny sizes were small enough that both approaches were computationally feasible.
The computational cost of the forward-equivalent simulation scales with the size of the sub-sampled rather than the full phylogeny.
Thus, in principle, it allows for simulation of phylogenies subsampled from arbitrarily large populations.

As an example, we consider a multi-type model with two \phenotypes $\Unfit$ and $\Fit$ with birth rates $\lambda_{\Unfit} = 0.25$ and $\lambda_{\Fit} = 1.0$,
death rates $\mu_{\Unfit} = 0.25$ and $\mu_{\Fit} = 0.5$,
and mutation rates $\gamma_{\Unfit,\Fit} = 0.1$ and $\gamma_{\Fit,\Unfit} = 0.25$,
respectively.
Thus, the $\Fit$ \phenotype has higher birth rates and death rates but a larger net speciation rate than the $\Unfit$ \phenotype, i.e., $\lambda_{\Fit} - \mu_{\Fit} > \lambda_{\Unfit} - \mu_{\Unfit}$.
Moreover, deleterious mutations occur at a higher rate than beneficial mutations.
We simulate the population for biological time $\tmax = 47$ and sample each lineage in the present population with probability $\rho = 10^{-9}$. 
With these simulation parameters, we find we often generate trees with leaf counts in the 10s to 100s.
This indicates that the full population size from which the phylogeny is sampled is in the 10s to 100s of \textit{billions}.
On a Macbook Pro with an Apple M1 Pro chip,
each tree in this simulation required just over a second of computation time on average.

\begin{figure}[b!]
\centering
\includegraphics[width=.8\linewidth]{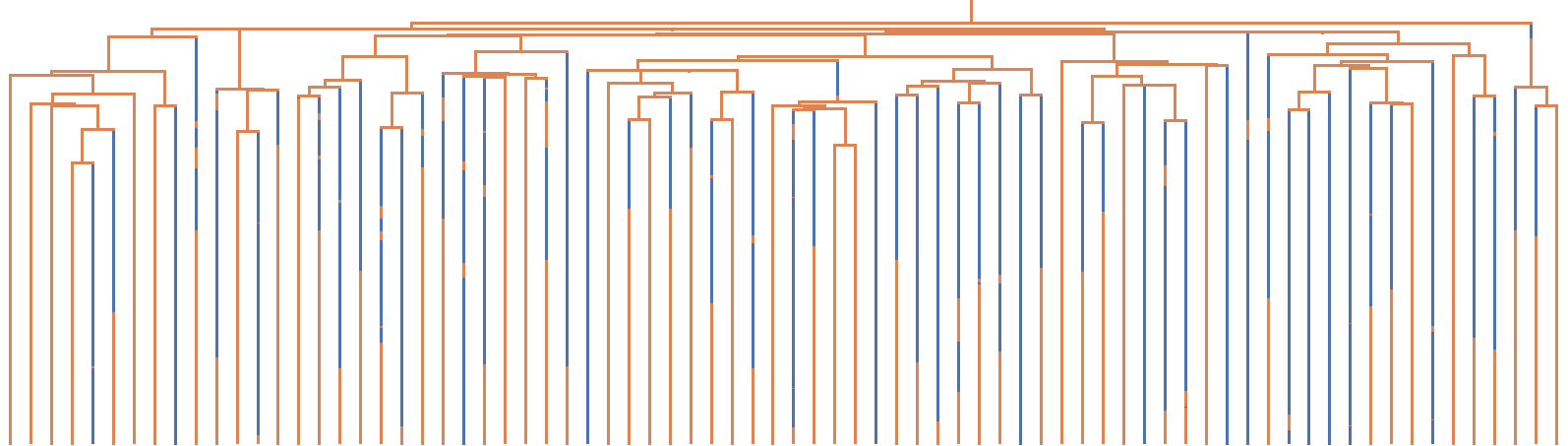}
\caption{A sub-sampled phylogeny from a multi-type model with population size in the billions. The $\Fit$ \phenotype is in orange and the $\Unfit$ \phenotype is in blue.}
\label{fig:multi-type-huge}
\end{figure}

In Figure \ref{fig:multi-type-huge},
we display a typical phylogeny drawn from this simulation.
From such a tree, we see several features that are relevant to understanding phylogenies generated from experiments involving billions of cells.
In particular, we observe that most birth events occur well before the present, and most (but not all) mutations occur after the most recent observed birth event.
Most lineages mutated to the $\Unfit$ \phenotype shortly after the most recent birth event, but mutated again to the $\Fit$ \phenotype before sampling at the present day.
Further information regarding observed mutation rates and birth rates, their time dependence, and the behavior of various approaches to estimating underlying process parameters, can also be investigated empirically using such simulated phylogenies.

\section{Discussion}

This paper has proposed a novel method for simulating reconstructed phylogenies that leads to substantial computational savings when there is sparse sampling or high levels of death in the population.
Many scientific studies rely on extensive phylogenetic simulation.
Unfortunately, when the full population size is in the billions or more, as is typical in epidemiological, cancer evolution, and antibody evolution applications,
the cost of the full simulation is prohibitive.
The forward-equivalent simulation we propose makes simulating the reconstructed phylogeny feasible in these domains when the set of sampled lineages is substantially smaller than the full population, enabling much more realistic simulations and benchmark studies.

We believe that the forward-equivalent simulation approach will facilitate improved inference of population dynamics.
We describe some directions for future work.

\begin{itemize}

    \item \textbf{Training of machine learning models.} Recently, deep learning approaches for estimating epidemiological parameters from reconstructed phylogenies have been proposed \cite{ThompsonLiebeskindScullyLandis2023}.
    These methods are often trained on large datasets of phylogenies simulated with different model parameters.
    By enabling faster simulation,
    our method can enable the training of such models.
    In particular, the forward-equivalent simulation can allow the simulated phylogenies to be faithful to realistic population sizes common in epidemiological, cancer evolution, and other domains.

    \item \textbf{Simulation at inference-time.} Some inferential methods require simulation at inference time. For example, ``Approximate Bayesian Computation'' involves computing summary statistics of observed phylogenies, and then stochastically searching for model parameters which, in simulation, generate phylogenies with similar summary statistics \cite{XieValentaEtienne2023}.
    A computational bottleneck in these approaches is the cost of simulation. We anticipate the forward-equivalent simulation may improve or enable simulation-based approaches to inference.

    \item \textbf{Identifiability.} Following the seminal paper of Louca and Pennell~\cite{Louca:2020aa}, there has been substantial interest in issues of identifiability in birth-death models \cite{legried2022class,legried2023identifiability}.
    These papers study single-type models with time-varying birth and death rates,
    and show that, due to sub-sampling, there are multiple values of the underlying process parameters that can induce the same distribution over observed phylogenies.
    This calls into question the ability to use phylogenies to reconstruct parameters driving diversification and extinction dynamics in the past.
    Our simulation approach is related to a lack of identifiability in multi-type models:
    indeed, the full model and the forward-equivalent model induce the same distribution over reconstructed phylogenies,
    so cannot be distinguished using the observed data alone.
    Thus, this paper establishes that the identifiability issues that occur in single-type models also occur in more general multi-type models,
    and we specify for each full model a different model that belongs to the same equivalence class.
    The current paper identifies the utility of this lack of identifiability for the purpose of simulation,
    but it will also be important to investigate the challenges this introduces for the purpose of inference.
    Moreover, a complete description of the classes of indistinguishable models and investigation of assumptions that may resolve identifiability issues, as in \cite{legried2022class,legried2023identifiability}, is left to future work.

\end{itemize}

\section{Materials and methods}
\label{sec:materials-methods}

\paragraph{Software implementation.}
We have made reproducible analysis available on GitHub (Code available at \url{https://github.com/songlab-cal/forward-equivalent-trees}).
We used the Python package BDMS \cite{BDMS}---which is developed by us and other collaborators---as a reference implementation of the full simulation, and developed wrapper code to implement the forward-equivalent mapping $\Theta \mapsto \Theta^\FE$ and perform simulation studies.
We use the Python package ETE \cite{10.1093/molbev/msw046} for tree manipulation and visualization.

\section*{Acknowledgments}
This research is supported in part by a Miller Research Fellowship (to M.C.), and a James S. McDonnell Foundation Postdoctoral Fellowship (to W.S.D), and an NIH grant R56-HG013117 (to Y.S.S).

\clearpage

\beginsupplement

\section*{Appendix}
\appendix

\renewcommand{\theequation}{S.\arabic{equation}}

\section{Further implementation details}
\label{sec:further-implementation}

In this section, we describe additional implementation details to enable Algorithm \ref{alg:full} to run in time which scales linearly in the number of events that occur in the full population.
First, we must implement G{\footnotesize ET}N{\footnotesize EXT}E{\footnotesize VENT} in constant time in the population size,
which is described in Appendix~\ref{app:get-next-event}.
We must also store, at any given time, the active population in a data structure that supports insertion, deletion, and selection of an element uniformly at random, in constant time in the population size,
which is described in Appendix~\ref{app:randomized-set}.
Finally, our ability to implement G{\footnotesize ET}N{\footnotesize EXT}E{\footnotesize VENT} requires we perform precomputations before simulation time.
These are described in Appendix~\ref{app:precomputation}.

\subsection{Implementing G{\footnotesize ET}N{\footnotesize EXT}E{\footnotesize VENT} in constant time in the population size}
\label{app:get-next-event}

As described in Section~\ref{sec:full-sim},
in order for Algorithm~\ref{alg:full} to have time complexity linear in the number of events in the full phylogeny,
we must implement the function G{\footnotesize ET}N{\footnotesize EXT}E{\footnotesize VENT} in constant time in the population size.
Our implementation relies on the following fact.
Consider a Poisson point process with time-varying rate given by $\theta(\tau)\geq 0$.
Let 
\begin{equation}
    \Theta(\tau,\tau')
        :=
        \int_{\tau}^{\tau'}
        \theta(\tau)\de \tau,
    \qquad
    \Theta^{-1}(\tau,z)
        :=
        \sup\{\tau'\in \R\mid \Theta(\tau,\tau')\leq z\}.
\end{equation}
Consider $n$ iid copies of this Poisson point process.
For a fixed time $\tau$,
consider the next time after $\tau$ that an event occurs in any one of these $n$ processes.
It is routine to show that this arrival time has distribution given by $\Theta^{-1}(\tau,Z/n)$ where $Z \sim \Exp(1)$ has exponential distribution with rate parameter $1$ (that is, has density given by $e^{-z}$ for $z \geq 0$).
Note that $\Theta(\tau,Z/n)$ can be infinite if $\lim_{\tau' \rightarrow \infty} \Theta(\tau,\tau') < \infty$ and $Z/n \geq \lim_{\tau' \rightarrow \infty} \Theta(\tau,\tau')$.
This corresponds to the event that none of the $n$ Poisson processes have any events occur after time $\tau$,
which can occur when the Poisson rate function is integrable.

The G{\footnotesize ET}N{\footnotesize EXT}E{\footnotesize VENT} function takes in the current time $\tau$ and the number of extant lineages of each \phenotype $|S_a|$.
Given the discussion in the previous paragraph,
this provides a way of determining when the next event in the population occurs,
and what type it is,
where the type specifies whether it is a birth, death, mutation, or sampling event,
and the \phenotypes of the parent and child nodes in the phylogeny.
For each event type,
we generate the time of the next event of that type.
The next event type is chosen to be the one that happens the soonest among these generated times.
Formally, for $a,b = 1,\dots,d$,
define
\begin{equation}
\label{eq:poisson-integrals}
\begin{gathered}
    \Lambda_{a,b}(\tau_1,\tau_2)
        = \int_{\tau_1}^{\tau_2} \lambda_{a,b}(\tau) \de \tau,
    \qquad
    M_a(\tau_1,\tau_2) = \int_{\tau_1}^{\tau_2} \mu_a(\tau) \de \tau,
    \\
    \Gamma_{a,b}(\tau_1,\tau_2)
        = \int_{\tau_1}^{\tau_2} \gamma_{a,b}(\tau) \de \tau,
    \qquad  
    \Psi_a(\tau_1,\tau_2)
        = \int_{\tau_1}^{\tau_2} \psi_a(\tau) \de \tau.
\end{gathered}
\end{equation}
Then define
\begin{equation}
\begin{gathered}
    \Lambda_{a,b}^{-1}(\tau,z)
        = \sup \{\tau' \in \reals \mid \Lambda_{a,b}(\tau,\tau') \leq z \},
    \qquad
    M_a^{-1}(\tau_1,z)
        = \sup \{\tau' \in \reals \mid M_a(\tau,\tau') \leq z \}
    \\
    \Gamma_{a,b}^{-1}(\tau_1,z)
        = \sup \{\tau' \in \reals \mid \Gamma_{a,b}(\tau,\tau') \leq z \}
    \qquad  
    \Psi_a^{-1}(\tau_1,z)
        = \sup \{\tau' \in \reals \mid \Psi_a(\tau,\tau') \leq z \},
\end{gathered}
\end{equation}
where these can be possibly infinite (for example, if one of the rate parameters is 0).
We draw exponential random variables $\{Z_{a,b}^{\gamma}\}_{a,b=1}^d$, $\{Z_a^\mu\}_{a=1}^d$, $\{Z_{a,b}^\gamma\}_{a,b=1}^d$, $\{Z_a^\psi\}_{a=1}^d$ iid from $\Exp(1)$ so that we have one exponential random variable for each event type.
The occurrence of the next birth event $a \rightarrow a,b$ occurs at time $\Lambda_{a,b}^{-1}(\tau,Z_{a,b}^\lambda/|S_a|)$;
the next death of a lineage with \phenotype $a$ occurs at time $M_a^{-1}(\tau,Z_a^\mu/|S_a|)$;
the next anagenetic mutation $a \rightarrow b$ occurs at time $\Gamma_{a,b}^{-1}(\tau,Z_{a,b}^\gamma/|S_a|)$;
and the next sampling event occurs at $\Psi_a^{-1}(\tau,Z_a^\psi/|S_a|)$.
The next event of any type occurs at the minimum of these,
and the type of the next event is determined by which process achieves the minimum.
Thus, G{\footnotesize ET}N{\footnotesize EXT}E{\footnotesize VENT} can be implemented by the pseudo-code in Algorithm \ref{alg:get-next-event}.
\begin{algorithm}[hbt!]
\caption{$\text{G{\footnotesize ET}N{\footnotesize EXT}E{\footnotesize VENT}}(\{|S_a|\},\tau)$}\label{alg:get-next-event}
\begin{algorithmic}
\State Draw $\{Z_{a,b}^{\gamma}\}_{a,b =1}^d$, $\{Z_a^\mu\}_{a=1}^d$, $\{Z_{a,b}^\gamma\}_{a,b=1}^d$, $\{Z_a^\psi\}_{a=1}^d$ iid from $\Exp(1)$;
\State $\tau_* \gets \min_{a,b}
        \big\{
            \Lambda_{a,b}^{-1}(\tau,Z_{a,b}^\lambda/|S_a|),
            M_a^{-1}(\tau,Z_a^\mu/|S_a|),
            \Gamma_{a,b}^{-1}(\tau,Z_{a,b}^\gamma/|S_a|),
            \Psi_a^{-1}(\tau,Z_a^\psi/|S_a|)
        \big\}$;
\State $\EventType \gets \arg\min
        \big\{
            \Lambda_{a,b}^{-1}(\tau,Z_{a,b}^\lambda/|S_a|),
            M_a^{-1}(\tau,Z_a^\mu/|S_a|),
            \Gamma_{a,b}^{-1}(\tau,Z_{a,b}^\gamma/|S_a|),
            \Psi_a^{-1}(\tau,Z_a^\psi/|S_a|)
        \big\}$;
\If{$\tau^* < 0$}
\Return $(0,\EventType)$;
\Else{
\Return $(\tau^*, \EventType)$;
}
\EndIf 
\end{algorithmic}
\end{algorithm}
Provided $\Lambda_{a,b}^{-1}(\tau,Z_{a,b}^\lambda/|S_a|)$, $M_a^{-1}(\tau,Z_a^\mu/|S_a|)$, $\Gamma_{a,b}^{-1}(\tau,Z_{a,b}^\gamma/|S_a|)$, $\Psi_a^{-1}(\tau,Z_a^\psi/|S_a|)$ can be evaluated in constant time,
the time complexity of Algorithm \ref{alg:get-next-event} is $O(d^2)$,
and in particular, does not grow with population size.

In the case that the time of the next event occurs after the present day,
we instead exit the loop and advance all extant lineages to the present day.
We then carry out the present day sampling process:
each lineage is sampled with probability $\rho_{a,0}$, and upon sampling, goes extinct with probability $r_{a,0}$.
If our model includes CSEs prior to the present day,
we follow a similar logic when the next event time occurs after the next CSE.
We can thus implement $\mathrm{GetNextEvent}$ using the pseudo-code in Algorithm \ref{alg:get-next-event} complemented by the additional CSE logic.
For simplicity, the pseudo-code represents a case without CSEs.

In some cases (for example, for constant rate Poisson processes),
$\Gamma_{a,b}^{-1}$, $M_a^{-1}$, $\Lambda_{a,b}^{-1}$, and $\Psi_a^{-1}$ have analytic forms that permit constant time evaluation.
When analytic forms are not available (which is typically the case for the forward-equivalent models),
we leverage precomputations.
Note that $\Theta^{-1}(\tau,z) = \Theta^{-1}(0,\Theta(0,\tau) + z)$.
Thus, to allow for constant time evaluation of $\Theta^{-1}$ at simulation time,
we can precompute $\Theta$ at an equispaced grid of $\tau \in [0,\tmax]$ and $\Theta^{-1}$ on an equispaced grid of $z \in [0,\Theta(0,\tmax)]$.
We describe the precomputations for $\Lambda_{a,b}^{-1}$, $M_a^{-1}(0,z)$, $\Gamma_{a,b}^{-1}(0,z)$, $\Psi_a^{-1}(0,z)$ in Appendix~\ref{app:precomputation}.
Then, at simulation time, we compute $\Theta(0,\tau)$ and $\Theta^{-1}(0,\Theta(0,\tau) + z)$ by linear interpolation with respect to this mesh.
This is a constant time operation, because, for example, the interval in which $\tau$ occurs has left and right indices (using Python indexing) given by $\lfloor \tau / \Delta \tau \rfloor$ and $\lceil \tau / \Delta \tau \rceil$,
where $\Delta \tau$ is the size of the gaps in the discrete grid.
Thus, the evaluation requires index-based array lookup.

We remark that in our code,\footnote{Code available at \url{https://github.com/songlab-cal/forward-equivalent-trees}}
the implementation of the mutation process is slightly different.
The event type specifies only the type of the \phenotype $a$ experiencing the mutation, not the \phenotype $b$ into which it mutates.
Conditional on \phenotype $a$ mutating, we draw \phenotype $b$ from the appropriate distribution.
In particular, rather than compute $\Gamma_{a,b}(\tau_1,\tau_2)$,
we compute
\begin{equation}
    \Gamma_a(\tau_1,\tau_2)
        = \int_{\tau_1}^{\tau_2} \sum_b \gamma_{a,b}(\tau) \de \tau,
\end{equation}
and define $\Gamma_a^{-1}(\tau,z)$ in the obvious way.
Then, conditional on a mutation occurring to \phenotype $a$ at time $\tau^*$, we draw $b$ from $\P(b) = \gamma_{a,b}(\tau^*) / \sum_{b'} \gamma_{a,b'}(\tau^*)$.
We do this based on the API of the BDMS simulation engine we use \cite{BDMS},
but do not claim there is any intrinsic advantage or disadvantage of either approach relative to the other.

\subsection{Randomized Set}
\label{app:randomized-set}

Algorithm \ref{alg:full} requires keeping sets of pointers to the active nodes in the population of each \phenotype at each point in time $\tau$.
This data structure must support insertion and deletion (see, for example, line \ref{line:remove-insert} of Algorithm \ref{alg:full}) as well as the ability to drawn an element of uniformly at random (see, for example, line \ref{line:get-random} of Algorithm \ref{alg:get-next-event}).
Such a data structure is called a ``randomized set,'' and is the data structure we use in our simulations.

\subsection{Precomputation}
\label{app:precomputation}

As described in Appendix~\ref{app:get-next-event},
in the absence of analytic forms,
we compute $\Theta^{-1}(\tau,z)$ by precomputing $\Theta(0,\tau)$ and $\Theta^{-1}(0,z)$ on a discrete grid, for $\Theta$ corresponding to $\Gamma_{a,b}$, $M_a$, $\Lambda_{a,b}$, or $\Psi_a$.
We here describe this precomputation for the forward-equivalent model.

For each \phenotype $a$,
we computed $E_a(\tau)$ according to \eqref{eq:reconstructed-E-ODE} using Euler's method on $[0,\tmax]$ with time discretization $\Delta \tau = .01$.
This gives $E_a(\tau)$ on a discrete grid, which we extend to any $\tau$ by linear interpolation.
The model parameters $(\pi^\FE,\allowbreak\lambda_{a,b}^\FE,\allowbreak\mu_a^\FE,\allowbreak\psi_a,r_a^\FE,\allowbreak\gamma_{a,b}^\FE,\allowbreak\rho_{a,l}^\FE,\allowbreak r_{a,l}^\FE)$ can then be precomputed on the same discrete grid using \eqref{eq:FE-params},
consuming space $O(d^2 \tmax / \Delta \tau)$
and requiring time $O(d^2\tmax/\Delta \tau)$.
We then precompute the integrals $\Gamma_{a,b}(0,\tau)$, $M_a(0,\tau)$, $\Lambda_{a,b}(0,\tau)$, $\Psi_a(0,\tau)$ on the same equispaced grid using numerical integration.
For example, this gives us $\Lambda_{a,b}(0,k\Delta \tau)$ for some finite sequence of integers $k = 0,1,\ldots K$.
This allows us to evaluate $\Lambda_{a,b}^{-1}(0,z)$ on the mesh $\{0,\Lambda_{a,b}(0,\Delta \tau),\ldots,\Lambda_{a,b}(0,K\Delta \tau)\}$.
This mesh is not equispaced.
We get an evaluation of $\Lambda_{a,b}^{-1}(0,z)$ on an equispaced grid of $[0,\Lambda_{a,b}(0,K\Delta\tau)]$ by linear interpolation of the values computed on the mesh $\{0,\Lambda_{a,b}(0,1\Delta \tau),\ldots,\Lambda_{a,b}(0,K\Delta \tau)\}$.
At simulation time, all evaluations are done using linear interpolation on these grids,
as described in Appendix~\ref{app:get-next-event}.

This approach requires storing precomputations in arrays whose memory footprint is\break $O(d^2 \tmax / \Delta \tau)$.
The time-complexity of the precomputation is dominated by the numerical integration of \eqref{eq:reconstructed-E-ODE},
whose time complexity using Euler's method is $O(d^2 \tmax / \Delta \tau)$.
The precomputation is performed only once, 
so this cost is amortized over the number of simulations carried out.
For the simulations in this paper,
the precomputations occurred in a few microseconds.
For very large \phenotype spaces (i.e., large $d$),
it is possible that both the space and time complexity of these precomputations would become prohibitive.

\subsection{Implementation based on a priority queue}
\label{sec:priority-queue}

An alternative simulation approach to that provided by Algorithm \ref{alg:full} has runtime $O(\#\text{nodes}(\cT) (d + \log(\#\text{nodes}(\cT))))$.
Thus, it may be faster when the \phenotype space is large. We briefly outline this approach here.

We again simulate forward in time $\tau$ and maintain the population of extant individuals at time $\tau$.
Additionally, we maintain a priority queue containing the next event for each individual in the extant population.
Thus, the next event in the population is given by the event in the priority queue with smallest time, which can be obtained in logarithmic time in the population size by virtue of the priority queue data structure.
The algorithm thus proceeds by advancing time to this event, and then performing the event.
When this event creates or modifies individuals (such as upon a birth event or a mutation event), the updated or newly created individuals are added to the population of extant individuals, and their next events are immediately sampled and pushed into the priority queue.

Inserting and popping from a priority queue has time complexity which grows logarithmically with the number of elements in the priority queue, which in this case corresponds to the extant population size.
For any fixed individual, generating the next event occurring to this individual has time complexity $O(|\mathcal{A}|)$.
This is due to generating the time and type of the next mutation event, which involves generating a random \phenotype from a state space of $|\mathcal{A}|$ \phenotypes with non-uniform probabilities stored in some precomputed arrays.
Thus, each iteration has time complexity $O(|\mathcal{A}| + \log(\#\text{nodes}(\cT)))$, and the number of iterations is equivalent to the number of events in the full phylogeny.
The total time complexity is thus $O(\#\text{nodes}(\cT)(|\mathcal{A}| +\log(\#\text{nodes}(\cT))))$.
This should be compared with the time complexity $O(\#\text{nodes}(\cT)|\mathcal{A}|^2)$ of the implementation in Algorithm \ref{alg:full}.
The alternative implementation with a priority queue has worse complexity in terms of the number of events in the full phylogeny but better complexity in terms of the size of the \phenotype space.
In applications with large \phenotype spaces, the priority queue based implementation may be preferred.
However, since our original algorithm has runtime that is strictly linear in the size of the full phylogeny, we focus on it for this work.

\section{Proof of Theorem~\ref{thm:forward-equivalence}\label{sec:FE-proof}}

Consider a non-empty phylogeny $\cT \ne T_\emptyset$.
Let $p(\cT)$ be the probability density of observing $\cT$ in the full model,
$p(\cT \mid \cT \ne T_\emptyset)$ be the probability density of observing $\cT$ in the full model conditional on the non-emptiness of $\cT$,
and let $p^\FE(\cT)$ be the probability density of observing $\cT$ in the forward-equivalent model.
We will show that $p(\cT \mid \cT \ne T_\emptyset)$ and $p^\FE(\cT)$ can both be computed by solving an ordinary differential equation.
We will show $p(\cT \mid \cT \ne T_\emptyset) = p^\FE(\cT)$ by showing that these differential equations are the same.
We draw on a large body of work which computes likelihoods in BDMS models using ordinary differential equations \cite{macPhersonLoucaMcLauglinJoyPennell2021}.

\subsection{The non-observation probability}
\label{app:E-ODE}

Eq.~\eqref{eq:reconstructed-E-ODE} agrees with (15) of \cite{macPhersonLoucaMcLauglinJoyPennell2021} together with the modifications described in their appendix for handling concerted sampling events.
Similar equations have appeared multiple time in earlier works; see \cite{macPhersonLoucaMcLauglinJoyPennell2021} for references.
For completeness, this section describes the derivation of these equations.
\paragraph{Between CSEs:} For $\tau$ in an open interval between CSEs, consider a lineage of \phenotype $a$ at time $\tau + \de \tau$.
    There are four ways that this lineage has no sampled descendants.
    We list this ways, and the probability of each, below.
    \begin{enumerate}

        \item No events occur in the interval $[\tau, \tau + \de \tau]$ and the lineage at time $\tau$ has no sampled descendants
        \begin{equation}
            \text{with probability}\;\;\Big(
                1 - \Big(\sum_{b=1}^d \lambda_{a,b}(\tau) + \mu_a(\tau) + \psi_a(\tau)  + \sum_{b=1}^d \gamma_{a,b}(\tau)\Big) \de \tau
            \Big)
            E_a(\tau).
        \end{equation}

        \item Birth $a \rightarrow (a,b)$ occurs in the interval $[\tau, \tau + \de \tau]$, and neither child has a sampled descendant
        \begin{equation}
            \text{with probability}\;\;\lambda_{a,b}(\tau)\de \tau E_a(\tau)E_b(\tau).
        \end{equation}

        \item Death occurs in the interval $[\tau, \tau + \de \tau]$
        \begin{equation}
            \text{with probability}\;\;\mu_a(\tau)\de \tau.
        \end{equation}

        \item Anagenetic mutation $a \rightarrow b$ occurs in the interval $[\tau, \tau + \de \tau]$,
        and the new lineage has not sampled descendants
        \begin{equation}
            \text{with probability}\;\;\gamma_{a,b}(\tau)\de \tau E_b(\tau).
        \end{equation}

    \end{enumerate}
    Summing these possibilities (and noting that items 2 and 4 can occur for any $b=1,\dots,d$) and rearranging gives the expression for $\frac{\de E_a(\tau)}{\de \tau}$ in \eqref{eq:reconstructed-E-ODE}.

\paragraph{Initialization:} $0^-$ refers to the moment immediately after that CSE occurring at the present,
    after that sampling and death events for that CSE have occurred.
    There is no possibility for any further sampling, so $E_a(0^-) = 1$.

\paragraph{At CSEs:} The equation $E_a(t_l^-) = \lim_{\tau^\uparrow t_l} E_a(\tau)$ holds because the time $t_l^-$ is after the sampling and death events for the CSE have occurred, so that $E_a(t_l^-)$ is the continuous extension of the extinction probability on the interval of time following the CSE.
    The time $t_l^+$ is immediately before the sampling and death events for the CSE have occurred. At this point in time, there is only one way in which the lineage can have no sampled descendants:
    the lineage is not sampled during the CSE (which has probability $1 - \rho_{a,l}$),
    and the lineage following the CSE has no sampled descendants (which has probability $E_a(t_l^-)$).
    Thus, $E_a(t_l^+) = (1-\rho_{a,l})E_a(t_l^-)$.

\subsection{Unconditional probability density of the reconstructed phylogeny}

Next, we describe how to compute $p(\cT)$ in the full model by solving a system of ordinary differential equations.
This argument is provided in \cite{macPhersonLoucaMcLauglinJoyPennell2021},
which itself unifies arguments appearing many times previously.
For completeness, we describe the argument here.

First, let us introduce some notation borrowed from \cite{macPhersonLoucaMcLauglinJoyPennell2021}.
The phylogeny $\cT$ consists of a collection of edges $\cE$,
where an edge on the tree is defined as the segment spanning two adjacent events, where events are either birth events, anagenetic mutation events, sampling events at a non-CSEs, or CSEs (with the accompany sampling or death events).
In the reconstructed phylogeny,
death events are simply sampling events after which no further descendant lineages were sampled.
Thus, sampling events that were followed by immediate death (whose probability is given by $\psi_a(\tau)$ and $r_a(\tau)$), 
and sampling events followed by survival but the failure to sample any future descendants,
look the same.
In the reconstructed phylogeny, sampling events can be internal to the tree or at leaf nodes, but all leaf nodes correspond to sampling events.

Each edge $e \in \cE$ spans a time interval $s_e < \tau < t_e$.
Letting $a_e$ denote the type of the lineage along edge $e$,
we let $g_e(\tau)$ denote the probability density that a lineage initialized at time $\tau$ with type $a_e$ would give rise to the sub-tree of $\cT$ ``anchored'' at time $\tau$ on edge $e$.
We use this terminology to refer to the sub-tree whose root is at time $\tau$ on edge $e$ and which contains all points on $\tau$ which are descendants of this point.
The probability density $g_e(\tau)$ \emph{does not} condition on non-emptiness.

The quantities $g_e(\tau)$ are given by the solution to an ODE which is integrated backwards in time along each edge, with the initial condition at $s_e$ determined by the solution along the more recent adjacent edges.

\paragraph{ODE:} The ODE is given by 
    \begin{equation}
    \label{eq:reconstructed-g-ODE}
    \begin{aligned}
        \frac{\de g_e(\tau)}{\de \tau}
            &=
            \lambda_{a,a}(\tau)\big(2E_a(\tau)-1\big)g_e(\tau) + \sum_{b\neq a} \lambda_{a,b}(\tau) \big(E_b(\tau)-1\big)g_e(\tau)
        \\
            &\qquad+
            \mu_a(\tau)\big(0-g_e(\tau)\big)
        \\
            &\qquad+
            \psi_a(\tau)\big(0-g_e(\tau)\big)
        \\
            &\qquad+
            \sum_{b\neq a} \gamma_{a,b}(\tau)(0-g_e(\tau)).
    \end{aligned}
    \end{equation}
    To see this, consider the probability density of the sub-tree anchored at time $\tau + \de \tau$ on edge $e$.
    The probability of this tree is given by the product of the probability that a lineage of type $a$ at time $\tau + \de \tau$ does not experience any observed events in the time interval $[\tau,\tau+\de \tau]$ and the probability density of the sub-tree anchored at time $\tau$.
    There are three ways that no events are observed in the time interval $[\tau,\tau+\de \tau]$.
    The first way is that no events occur.
    This has probability $1 - \Big(\sum_{b=1}^d \lambda_{a,b}(\tau) + \mu_a(\tau) + \psi_a(\tau) + \sum_{b \neq a}\gamma_{a,b}(\tau)\Big) \de \tau$.
    The second is that a birth of type $a \rightarrow a,a$ occurs and one of the child lineages has no sampled descendants.
    This has probability $2\lambda_{a,a}(\tau)\de \tau E_a(\tau)$,
    where the factor of $2$ results from the possibility that either lineage is the one without sampled descendants.
    The third way is that a birth $a\rightarrow (a,b)$ for $b \neq a$ occurs and the child lineage of type $b$ has no sampled descendants.
    This has probability $\lambda_{a,b}(\tau)\de \tau E_b(\tau)$.
    Thus,
    \begin{align}
        g_e(\tau + \de \tau)
            =
            \Bigg\{
                1 - & \Bigg[\sum_{b=1}^d \lambda_{a,b}(\tau) + \mu_a(\tau) + \psi_a(\tau) + \sum_{b \neq a}\gamma_{a,b}(\tau)\Bigg] \de \tau
                 \\ \nonumber
                &         +
                2 \lambda_{a,a}(\tau) \de \tau E_a(\tau) +
                \sum_{b \neq a} \lambda_{a,b}(\tau)\de \tau E_b(\tau)
            \Bigg\}
            g_e(\tau).
    \end{align}
    This rearranges to \eqref{eq:reconstructed-g-ODE}.

\paragraph{Initial conditions:} The initial condition at $s_e$ depends on the type of event occurring at the terminal node of an edge.

    \begin{itemize}
    \begin{subequations}

        \item \textbf{Births:} For birth events resulting in edges $(e_1,e_2)$ with types $(a,b)$ or $(b,a)$,

        \label{eq:reconstructed-g-init}
        \begin{equation}
        \label{eq:reconstructed-g-init-birth}
            g_e(s_e)
                =
                \lambda_{a,b}(s_e)g_{e_1}(s_e)g_{e_2}(s_e).
        \end{equation}
        The factor $\lambda_{a,b}(s_e)$ gives the probability density of a birth occurring at time $\tau$, and $g_{e_1}(s_e)g_{e_2}(s_e)$ the probability of the two children giving rise to their observed sub-trees.

        \item \textbf{Anagenetic mutations:} For anagenetic mutations $a \rightarrow b$ with offspring edge $e_1$,
        \begin{equation}
        \label{eq:reconstructed-g-init-mut}
            g_e(s_e)
                =
                \big(\gamma_{a,b}(s_e) + \lambda_{a,b}(s_e)E_a(s_e)\big)g_{e_1}(s_e).
        \end{equation}
        An anagenetic mutation $a \rightarrow b$ in the reconstructed phylogeny can
        result from two types of events in the full phylogeny.
        The first type of event is the anagenetic mutation $a \rightarrow b$ in the full phylogeny with the child lineage giving rise to the observed sub-tree.
        This has probability density $\gamma_{a,b}(s_e)g_{e_1}(s_e)$.
        The second type of event is a birth $a \rightarrow (a,b)$,
        with the child of type $a$ having no observed descendants and the child of type $b$ giving rise to the observed sub-tree.
        This has probability density $\lambda_{a,b}(s_e)E_a(s_e)g_{e_1}(s_e)$.
        Adding these two terms gives \eqref{eq:reconstructed-g-init-mut}.

        \item \textbf{Non-CSE sampling events:} These are initialized by
        \begin{equation}
        \label{eq:reconstructed-g-init-sampling}
            g_e(s_e)
                =
                \begin{cases}
                    \psi_a(s_e)(r_a(s_e) + (1-r_a(s_e))E_a(s_e))
                    \quad& \text{for leaf nodes},
                    \\
                    \psi_a(s_e)(1-r_a(s_e))g_{e_1}(s_e)
                    \quad& \text{for internal nodes}.
                \end{cases}
        \end{equation}
        A sampled leaf node at a non-CSE time can result from two types of events in the full phylogeny.
        The first type is that the lineage was sampled and immediately went extinct.
        This has probability density $\psi_a(s_e)r_a(s_e)$.
        The second type is that the lineage was sampled, did not immediately go extinct, and then had no sampled descendants.
        This has probability density $\psi_a(s_e)(1-r_a(s_e))E_a(s_e)$.
        Adding these two terms gives the first line of \eqref{eq:reconstructed-g-init-sampling}.
        
        A sampled internal node at a non-CSE time can result from only one type of event in the full phylogeny.
        This event is that the lineage was sampled, did not immediately go extinct, and then gave rise to the observed sub-tree.
        This has probability density $\psi_a(s_e)(1-r_a(s_e))g_{e_1}(s_e)$,
        giving the second line of \eqref{eq:reconstructed-g-init-sampling}.

        \item \textbf{CSE sampling events:}
        These are initialized by
        \begin{equation}
        \label{eq:reconstructed-g-init-CSE}
            g_e(t_l^+)
                =
                \begin{cases}
                    (1-\rho_{a,l})g_{e_1}(t_l^-)
                    \quad& \text{for non-sampled nodes},
                    \\
                    \rho_{a,l}(r_{a,l} + (1-r_{a,l})E_a(t_l^-))
                    \quad& \text{for sampled leaf nodes},
                    \\
                    \rho_{a,l}(1-r_{a,l})g_{e_1}(t_l^-)
                    \quad& \text{for sampled internal nodes}.
                \end{cases}
        \end{equation}
        If no sampling occurs at the CSE in the reconstructed phylogeny,
        then no sampling occurred in the full phylogeny and the lineage at time $t_l^-$ gave rise to the observed sub-tree.
        This has probability density $(1-\rho_{a,l})g_{e_1}(t_l^-)$,
        giving the first line of \eqref{eq:reconstructed-g-init-CSE}.
        The remaining two lines of \eqref{eq:reconstructed-g-init-CSE} follow by exactly the same argument leading to the two lines of \eqref{eq:reconstructed-g-init-sampling},
        with $\psi_a(s_e)$ replaced by $\rho_{a,l}$ and $r_a(s_e)$ replaced by $r_{a,l}$.
        
    \end{subequations}
    \end{itemize}

One initializes these ODEs at every leaf node in the reconstructed phylogeny, and then integrates backwards in time along all lineages until reaching the root.
The value $g_e(\tmax)$ on the root edge of the tree is the probability density of observing $\cT$.

\subsection{Probability density conditional on non-emptiness}

For an edge $e$ and time $\tau$ along edge $e$ in $\cT$,
let $\gbar_e(\tau)$ denote the probability density that a lineage initialized at time $\tau$ with type $a_e$ gives rise to the sub-tree of $\cT$ anchored at time $\tau$ on edge $e$,
conditional on such a lineage having an observed descendant.
That is, according to Bayes rule,
\begin{equation}
\label{eq:gbar-bayes}
    \gbar_e(\tau)
        =
        \frac{g_e(\tau)}{1 - E_a(\tau)}.
\end{equation}
Let $g_e^\FE(\tau)$ denote the quantity $g_e(\tau)$ for the forward equivalent model.
That is, $g_e^\FE(\tau)$ is defined using parameters $(\pi^\FE,\allowbreak\lambda_{a,b}^\FE,\allowbreak\mu_a^\FE,\allowbreak\psi_a^\FE,r_a^\FE,\allowbreak\gamma_{a,b}^\FE,\allowbreak\rho_{a,l}^\FE,\allowbreak r_{a,l}^\FE)$ in \eqref{eq:reconstructed-g-ODE} and \eqref{eq:reconstructed-g-init}, with $E_a^{\FE}(\tau)$ as determined by these parameters according to Appendix~\ref{app:E-ODE}.
Denoting by $e_*$ the root lineage of $\cT$,
Theorem \ref{thm:forward-equivalence} will follow by establishing
\begin{equation}
\label{eq:cond-density-is-FE-density}
    \gbar_{e_*}(\tmax) = g_{e_*}^\FE(\tmax).
\end{equation}
The remainder of the proof consists of establishing \eqref{eq:cond-density-is-FE-density}.
This is done by showing that $\gbar_e(\tau)$ obeys the same ODEs and initial conditions as $g_e^\FE(\tau)$,
given by \eqref{eq:reconstructed-g-ODE} and \eqref{eq:reconstructed-g-init} with the forward-equivalent model parameters.

First, we collect some facts about the forward-equivalent model parameters.
By \eqref{eq:FE-params},
$r_{a,0}^\FE = r_{a,0} + (1-r_{a,0})E_a(0^-) = 1$ and $\rho_{a,0}^\FE = \rho_{a,0}/(1-E_a(0^+)) = 1$.
As explained in Appendix~\ref{app:E-ODE}, $E_a^\FE(0^-) = 1$.
Thus, $E_a^\FE(0^+) = (1 - \rho_{a,0}^\FE)E_a^\FE(0^-) = 0$.
Then, because $\mu_a^\FE(0) = 0$, 
Eqs.~\eqref{eq:reconstructed-E-ODE} and \eqref{eq:reconstructed-E-ODE} imply $E_a^\FE(\tau) = 0 $ for all $\tau \geq 0$ (excluding $E_a^\FE(0^-)$).
We collect these identities in the next display:
\begin{equation}
    \rho_{a,0}^\FE = 1,
    \quad
    \rho_{a,0}^\FE = 1,
    \quad
    E_a^\FE(0^-) = 1,
    \quad
    E_a^\FE(\tau) = 0 \text{ for $\tau > 0^-$}.
\end{equation}

\noindent \textbf{Agreement of ODEs.}
We differentiate \eqref{eq:gbar-bayes} to get the ODE satisfied by $\gbar_e(\tau)$.
In particular,
\begin{equation}
    \frac{\de \gbar_e(\tau)}{\de \tau}
        =
        \frac{\frac{\de}{\de \tau}g_e(\tau)}{1 - E_a(\tau)}
        +
        \frac{g_e(\tau)}{(1-E_a(\tau))^2} \frac{\de E_a(\tau)}{\de \tau}.
\end{equation}
Using \eqref{eq:reconstructed-E-ODE} and \eqref{eq:reconstructed-g-ODE},
this is
\begin{equation}
\begin{aligned}
    =
    &\frac{(\lambda_{a,a}(\tau)E_a(\tau) + \sum_b\lambda_{a,b}(\tau)(E_b(\tau)-1)-\mu_a(\tau) - \psi_a(\tau) - \sum_b\gamma_{a,b}(\tau))g_e(\tau)}{1 - E_a(\tau)}
    \\
        &\quad+
        \frac{g_e(\tau)}{(1-E_a(\tau))^2}
        \Big(
            \sum_b\lambda_{a,b}(\tau)(E_a(\tau)E_b(\tau) - E_a(\tau))
            +
            \mu_a(\tau)(1 - E_a(\tau))
    \\
            &\qquad\qquad\qquad+
            \psi_a(\tau)(0-E_a(\tau))
            +
            \sum_b \gamma_{a,b}(\tau)(E_b(\tau) - E_a(\tau))
        \Big).
\end{aligned}
\end{equation}
The terms involving $\lambda_{a,b}(\tau)$ are $g_e(\tau)/(1-E_a(\tau))$ multiplied by
\begin{equation}
\begin{aligned}
    &\lambda_{a,a}(\tau)E_a(\tau) + \sum_b \lambda_{a,b}(\tau)(E_b(\tau)-1)
        + 
        \frac{\sum_b\lambda_{a,b}(\tau)(E_a(\tau)E_b(\tau)-E_a(\tau))}{1 - E_a(\tau)}
    \\
        &\qquad=
        \lambda_{a,a}(\tau)E_a(\tau) - \sum_b \lambda_{a,b}^\FE(\tau)
        -
        \sum_b\frac{1-E_b(\tau)}{1-E_a(\tau)}E_a(\tau)\lambda_{a,b}(\tau).
\end{aligned}
\end{equation}
The terms involving $\mu_a(\tau)$ are $g_e(\tau)/(1-E_a(\tau))$ multiplied by
\begin{equation}
    -\mu_a(\tau) + \frac{\mu_a(\tau)(1-E_a(\tau))}{1-E_a(\tau)} = 0 = \mu_a^\FE(\tau).
\end{equation}
The terms involving $\psi_a(\tau)$ are $g_e(\tau)/(1-E_a(\tau))$ multiplied by
\begin{equation}
    -\psi_a(\tau) - \psi_a(\tau)\frac{E_a(\tau)}{1-E_a(\tau)}
    =
    -\psi_a^\FE(\tau).
\end{equation}
The terms involving $\gamma_{a,b}(\tau)$ are $g_e(\tau)/(1-E_a(\tau))$ multiplied by
\begin{equation}
    -\sum_b \gamma_{a,b}(\tau) + \frac{\sum_b \gamma_{a,b}(\tau)(E_b(\tau)-E_a(\tau))}{1 - E_a(\tau)}
        =
        -\sum_b \frac{1 - E_b(\tau)}{1-E_a(\tau)}\gamma_{a,b}(\tau).
\end{equation}
Combining these calculations,
\begin{equation}
\begin{aligned}
    &\frac{\de \gbar_e(\tau)}{\de \tau}
    \\
        &=
        \Big(
            -\sum_b \lambda_{a,b}^\FE(\tau)
            -\mu_a^\FE(\tau)
            -\psi_a^\FE(\tau)
            -\sum_b \frac{1-E_b(\tau)}{1-E_a(\tau)}(\gamma_{a,b}(\tau) + E_a(\tau)\lambda_{a,b}(\tau))
            +
            \lambda_{a,a}(\tau)E_a(\tau)
        \Big)\gbar_e(\tau)
    \\
        &=
        \Big(
            -\sum_b \lambda_{a,b}^\FE(\tau)
            -\mu_a^\FE(\tau)
            -\psi_a^\FE(\tau)
            -\sum_{b\neq a} \frac{1-E_b(\tau)}{1-E_a(\tau)}(\gamma_{a,b}(\tau) + E_a(\tau)\lambda_{a,b}(\tau))
        \Big)\gbar_e(\tau)
    \\
        &=
        \Big(
            -\sum_b \lambda_{a,b}^\FE(\tau)
            -\mu_a^\FE(\tau)
            -\psi_a^\FE(\tau)
            -\sum_{b\neq a} \gamma_{a,b}^\FE(\tau)
        \Big)\gbar_e(\tau).
\end{aligned}
\end{equation}
where the second inequality uses that $(1-E_a(\tau))/(1-E_a(\tau)) = 1$ and $\gamma_{a,a}(\tau) = 0$.
Because $E_a^\FE(\tau) = 0$ for all $\tau \geq 0$,
this ODE is \eqref{eq:reconstructed-g-ODE} under the forward-equivalent model parameters.
\\

\noindent \textbf{Agreement of initial conditions.}
We now check agreement of the initial conditions.
\begin{itemize}

    \item \textbf{Births:} For birth events resulting in edges $(e_1,e_2)$ with types $(a,b)$ or $(b,a)$,
    Bayes rule and \eqref{eq:reconstructed-g-init-birth} give
    \begin{equation}
    \begin{aligned}
        \gbar_e(s_e)
            &=
            \frac{g_e(s_e)}{1-E_a(s_e)}
            =
            \frac{\lambda_{a,b}(\tau)g_{e_1}(s_e)g_{e_2}(s_e)}{1-E_a(s_e)}
            =
            \lambda_{a,b}(\tau)(1-E_b(s_e))\frac{g_{e_1}(s_e)}{1-E_a(s_e)}\frac{g_{e_2}(s_e)}{1-E_b(s_e)}
        \\
            &=
            \lambda_{a,b}^\FE(\tau)\gbar_{e_1}(\tau)\gbar_{e_2}(\tau),
    \end{aligned}
    \end{equation}
    which is \eqref{eq:reconstructed-g-init-birth} for the forward-equivalent model.

    \item \textbf{Anagenetic mutations:} For anagenetic mutations $a \rightarrow b$ with offspring edge $e_1$,
    Bayes rule and \eqref{eq:reconstructed-g-init-mut} give
    \begin{equation}
    \begin{aligned}
        \gbar_e(s_e)
            &=
            \frac{g_e(s_e)}{1-E_a(s_e)}
            =
            \frac{\gamma_{a,b}(s_e)+\lambda_{a,b}(s_e)E_a(s_e)}{1 - E_a(s_e)}g_{e_1}(s_e)
        \\
            &=
            \frac{(1-E_b(s_e))(\gamma_{a,b}(s_e)+\lambda_{a,b}(s_e)E_a(s_e))}{1-E_a(s_e)}\frac{g_{e_1}(s_e)}{1-E_b(s_e)}
            = \gamma_{a,b}^\FE(s_e) \gbar_{e_1}(s_e),
    \end{aligned}
    \end{equation}
    which is \eqref{eq:reconstructed-g-init-mut} for the forward-equivalent model.

    \item \textbf{Non-CSE sampling events:}
    At leaf nodes,
    \begin{equation}
    \begin{aligned}
        \gbar_e(s_e)
            &=
            \frac{g_e(s_e)}{1-E_a(s_e)} 
            =
            \frac{\psi_a(s_e)(r_a(s_e)+(1-r_a(s_e))E_a(s_e))}{1-E_a(s_e)} 
        \\
            &=
            \psi_a^\FE(s_e)(r_a^\FE(s_e)+(1-r_a^\FE(s_e))E_a^\FE(s_e)),
    \end{aligned}
    \end{equation}
    which is the first line of \eqref{eq:reconstructed-g-init-sampling} for the forward-equivalent model.
    At internal nodes,
    \begin{equation}
    \begin{aligned}
        \gbar_e(s_e)
            &=
            \frac{g_e(s_e)}{1-E_a(s_e)} 
            =
            \frac{\psi_a(s_e)(1-r_a(s_e))g_{e_1}(s_e)}{1-E_a(s_e)} 
        \\
            &=
            \frac{\psi_a(s_e)}{1 - E_a(s_e)} (1-E_a(s_e))(1 - r_a(s_e)) \frac{g_{e_1}(s_e)}{1 - E_a(s_e)}
        \\
            &=
            \psi_a^\FE(s_e)(1 - r_a^\FE(s_e))g_{e_1}^\FE(s_e),
    \end{aligned}
    \end{equation}
    which is the second line of \eqref{eq:reconstructed-g-init-sampling} for the forward equivalent model.

    \item \textbf{CSE sampling events:} Using $\rho_{a,l}^\FE = \rho_{a,l}/(1-E_a(t_l^+))$ (by \eqref{eq:FE-params}) and $E_a(t_l^+) = (1-r_{a,l})E_a(t_l^-)$ (by \eqref{eq:reconstructed-E-ODE}),
    that $1 - \rho_{a,l} = \frac{1-\rho_{a,l}^\FE}{1-\rho_{a,l}^\FE E(t_l^-)}$.
    Then, for non-sampling at CSE times,
    \begin{equation}
    \begin{aligned}
        \gbar_e(t_l^+)
            &=
            \frac{g_e(t_l^+)}{1-E_a(t_l^+)} = \frac{(1-\rho_{a,l})g_{e_1}(t_l^-)}{1 - (1-\rho_{a,l})E_1(t_l^-)}
        \\
            &=
            \frac{\frac{1-\rho_{a,l}^\FE}{1-\rho_{a,l}^\FE E(t_l^-)}g_{e_1}(t_l^-)}{1 - \frac{1-\rho_{a,l}^\FE}{1-\rho_{a,l}^\FE E(t_l^-)}E_1(t_l^-)}
            =
            (1-\rho_{a,l}^\FE) \frac{g_{e_1}(t_l^-)}{1-E_1(t_l^-)} = (1-\rho_{a,l}^\FE) \gbar_{e_1}(t_l^-),
    \end{aligned}
    \end{equation}
    which is the first line of \eqref{eq:reconstructed-g-init-CSE} for the forward equivalent model.
    For sampled leaf nodes at CSE times,
    \begin{equation}
    \begin{aligned}
        \gbar_e(t_l^+)
            &=
            \frac{g_e(t_l^+)}{1-E_a(t_l^+)}
            =
            \frac{\rho_{a,l}(r_{a,l}+(1-r_{a,l})E_a(t_l^-))}{1-E_a(t_l^+)}
            =
            \rho_{a,l}^\FE r_{a,l}^\FE,
    \end{aligned}
    \end{equation}
    which is the second line of \eqref{eq:reconstructed-g-init-CSE} for the forward equivalent model (recall $E_a^\FE(t_l^+) = 0$).
    For sampled internal nodes at CSE times,
    \begin{equation}
    \begin{aligned}
        \gbar_e(t_l^+)
            &=
            \frac{g_e(t_l^+)}{1-E_a(t_l^+)}
            =
            \frac{\rho_{a,l}(1-r_{a,l})g_{e_1}(t_l^-)}{1-E_a(t_l^+)}
            =
            \rho_{a,l}^\FE(1-r_{a,l}^\FE) \frac{g_{e_1}(t_l^-)}{1 - E_a(t_l^-)}
            =
            \rho_{a,l}^\FE(1-r_{a,l}^\FE) \gbar_{e_1}(t_l^-).
    \end{aligned}
    \end{equation}
    where we have used $1 - r_{a,l}^\FE = (1-r_{a,l})(1-E_a(t_l^-))$ by \eqref{eq:FE-params}.
    This is the third line of \eqref{eq:reconstructed-g-init-CSE} for the forward equivalent model.

\end{itemize}

\noindent Thus, we have confirmed that $\gbar_e(\tau)$ and $g_e^\FE(\tau)$ obey the same ODE and initial conditions.
We conclude \eqref{eq:cond-density-is-FE-density},
which completes the proof of Theorem \ref{thm:forward-equivalence}.

\section{Proof of Proposition \ref{prop:general-speedup} and Proposition \ref{proposition:bd-speedup}}
\label{sec:speedup}

For a tree $T$, let $\#\text{nodes}(T)$ be the number of nodes of $T$.
In the context of the BDMS model, nodes correspond to (1) birth events, (2) death (or extinction) events, (3) mutation events, and (4) sampling events.
With suitable preprocessing, the runtime required to simulate a full tree $T$ from a given BDMS model is (up to a constant) exactly $\#\text{nodes}(T)$.
We mean `full' tree $T$, because the `observed' or `reconstructed' tree, denoted $r(T)$, is instead the sub-tree induced by the sampled leaves, and usually the desired output from the simulation.
Since a full tree might have an empty reconstructed tree, the time required to produce a non-empty reconstructed tree from the original BDMS model via retrying is $\#\text{nodes}(T_1) + \#\text{nodes}(T_2) + \dots + \#\text{nodes}(T_K)$ where $T_1, T_2, \dots$ is an infinite sequence of full trees simulated i.i.d. from the BDMS model and $T_K$ is the first full tree with non-empty reconstruction (so that $K$ is a stopping time random variable).
As we have shown, given a BDMS model, there exists an equivalent BDMS model---in the sense of providing the same distribution over reconstructed trees conditional on non-emptyness---such that under the new BDMS model there is no death ($\mu_a^\FE(\tau) = 0$) and all leaves are sampled.
In other words, the new model can simulate exactly from the original BDMS model \emph{without having to perform any retries nor reconstruction}.
The runtime of the forward-equivalent BDMS model (under the obvious coupling) is hence just $\#\text{nodes}(r(T_K))$ rather than $\#\text{nodes}(T_1) + \#\text{nodes}(T_2) + \dots + \#\text{nodes}(T_K)$, which may be orders of magnitudes faster if $r(T_K)$ is significantly smaller than $T_K$ or if a lot of retries are needed (i.e., if $K$ is large).
Let $\PP, \EE$ be probabilities and expectations under the original BDMS model, and let $\PP^\FE, \EE^\FE$ be probabilities and expectations under the forward-equivalent BDMS model.
The expected runtime of the original BDMS model is:
\[R = \EE[\#\text{nodes}(T_1) + \#\text{nodes}(T_2) + \dots + \#\text{nodes}(T_K)]\]
whereas the expected runtime of the forward-equivalent BDMS model is:
\[R^\FE = \EE[\#\text{nodes}(r(T_K))]\]
It is immediately clear that $R \ge R^\FE$, i.e., it is faster to simulate under the forward-equivalent model.
Our Proposition \ref{prop:general-speedup} gives a more precise characterization of this speedup, in a way that is easy to compute with time discretization and dynamic programming.
We now prove the Lemma.

\medskip\noindent
\textbf{Proof of Proposition \ref{prop:general-speedup}}.
Under our setup, the runtime of the forward-equivalent model is:
\begin{equation*}
R^\FE = \EE[\#\text{nodes}(r(T_K))] = \EE[\#\text{nodes}(r(T_1)) | r(T_1) \ne T_\emptyset]
\end{equation*}
As for the runtime of the original BDMS model, let $S = \#\text{nodes}(T_1) + \#\text{nodes}(T_2) + \dots + \#\text{nodes}(T_K)$.
By total expectation, we have that:
\begin{align*}
R =& \EE[S] \\
=& \PP(r(T_1) = T_\emptyset)\EE[S | r(T_1) = T_\emptyset] \\
&+ \PP(r(T_1) \ne T_\emptyset)\EE[S | r(T_1) \ne T_\emptyset] = \\
=& \PP(r(T_1) = T_\emptyset)(\EE[\#\text{nodes}(T_1) | r(T_1)= T_\emptyset] + R) \\
&+ \PP(r(T_1) \ne T_\emptyset)\EE[\#\text{nodes}(T_1) | r(T_1) \ne T_\emptyset] \\
=& \PP(r(T_1) = T_\emptyset) R + \EE[\#\text{nodes}(T_1)]
\end{align*}
\bigskip
so that solving for $R$ we get:
\begin{equation*}
R = \frac{\EE[\#\text{nodes}(T_1)]}{\PP(r(T_1) \ne T_\emptyset)}
\end{equation*}
Another way to show this is by noting that the runtime of tree $T_i$ should be considered if and only if all of $r(T_1), \dots, r(T_{i - 1})$ are empty.
Crucially, for fixed $i$, whether we consider the runtime from tree $T_i$ does not depend on whether $T_i$ is empty or not.
(Note that this is not true for $T_K$; the runtime of $T_K$ is biased by the fact that it is, by definition, non-empty.)
Hence,
\begin{align*}
R = \sum_{i = 1}^\infty \PP(r(T_1) = T_\emptyset)^{i - 1}\EE[\#\text{nodes}(T_i)] = \frac{\EE[\#\text{nodes}(T_1)]}{\PP(r(T_1) \ne T_\emptyset)}
\end{align*}
Combining our identities for $R$ and $R^\FE$ we obtain:
\begin{align*}
\frac{R}{R^\FE} = \frac{\EE[\#\text{nodes}(T_1)]}{\PP(r(T_1) \ne T_\emptyset)\EE[\#\text{nodes}(r(T_1)) | r(T_1) \ne T_\emptyset]}
\end{align*}
Since $\EE[\#\text{nodes}(r(T_1)) | r(T_1) = T_\emptyset] = 0$, we can complete the denominator using the law of total expectation to obtain:
\begin{align*}
\frac{R}{R^\FE} = \frac{\EE[\#\text{nodes}(T_1)]}{\EE[\#\text{nodes}(r(T_1))]}
\end{align*}
as claimed. \qed

With this, it is easy to prove Proposition \ref{proposition:bd-speedup}:

\medskip\noindent
\textbf{Proof of Proposition \ref{proposition:bd-speedup}.} We compute $\EE[\#\text{nodes}(T)]$ and $\EE[\#\text{nodes}(r(T))]$ and take the ratio, which by the lemma provides the speedup.
Note that
\[\EE[\#\text{nodes}(r(T))] = 
2 \rho \EE[\text{number of leaf nodes}]\]
and
\[\EE[\#\text{nodes}(T)] = 2 \EE[\text{number of leaf nodes}] + 2 \EE[\text{number of deaths}]\]
so it suffices to compute $\EE[\text{number of leaf nodes}]$ and $\EE[\text{number of deaths}]$, which we do separately.
Each of these quantities is given by an ODE; let:
\[l(t) = \EE[\text{number of leaf nodes} | \text{start with one individual at time $t$}]\]
and
\[d(t) = \EE[\text{number of deaths} | \text{start with one individual at time $t$}]\]
then:
\begin{align*}
\dot{l}(t) &= (\lambda - \mu)l(t);\quad l(0) = 1 \\
\dot{d}(t) &= \mu + (\lambda - \mu) d(t);\quad d(0) = 0
\end{align*}
If $\lambda = \mu$ then the solutions are $l(t) = 1, d(t) = t\mu$.
Otherwise:
\begin{align*}
l(t) &= e^{(\lambda - \mu)t} \\
d(t) &= \frac{\mu(e^{(\lambda - \mu)t} - 1)}{\lambda - \mu}
\end{align*}
Evaluating at $t = 1$ provides the desired quantities from which we can derive the given ratio.

\clearpage
\bibliographystyle{abbrv}
\bibliography{references}

\clearpage
\section*{Supplementary Figures}
\label{app:dist-test}

\begin{figure}[ht!]
\centering
\includegraphics{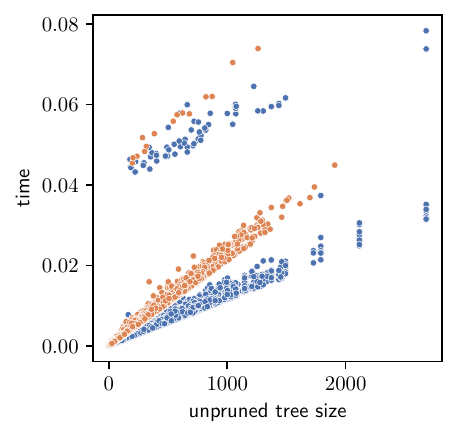}
\includegraphics{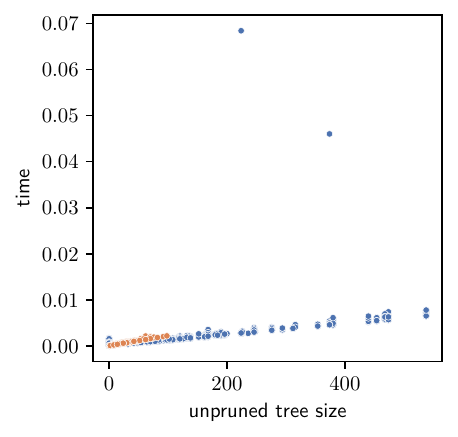}\\
\includegraphics{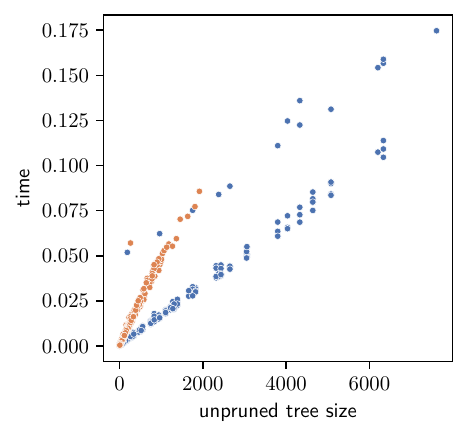}\\
\includegraphics{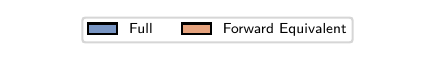}
\caption{Computation time 
required for each individual tree against the unpruned tree size across various simulation settings.
We see that, in correspondence with the discussion in Appendix~\ref{sec:further-implementation} and Section~\ref{sec:complexity},
the computation time is approximately linear in the size of the unpruned tree.
The forward-equivalent model is slower than the full simulation after controlling for the size of the unpruned tree due to the computational overhead of simulating from a BDMS model with time-varying rates.
The greater efficiency of the forward-equivalent simulation comes from simulating trees with substantially smaller unpruned tree size.
Top left: single-type model without death. Top right: single-type model with death. Bottom: multi-type model. Each point is a single tree.}
\label{fig:time-vs-unpruned-size}
\end{figure}

\begin{figure}[ht!]
\centering
\includegraphics[width=0.8\textwidth]{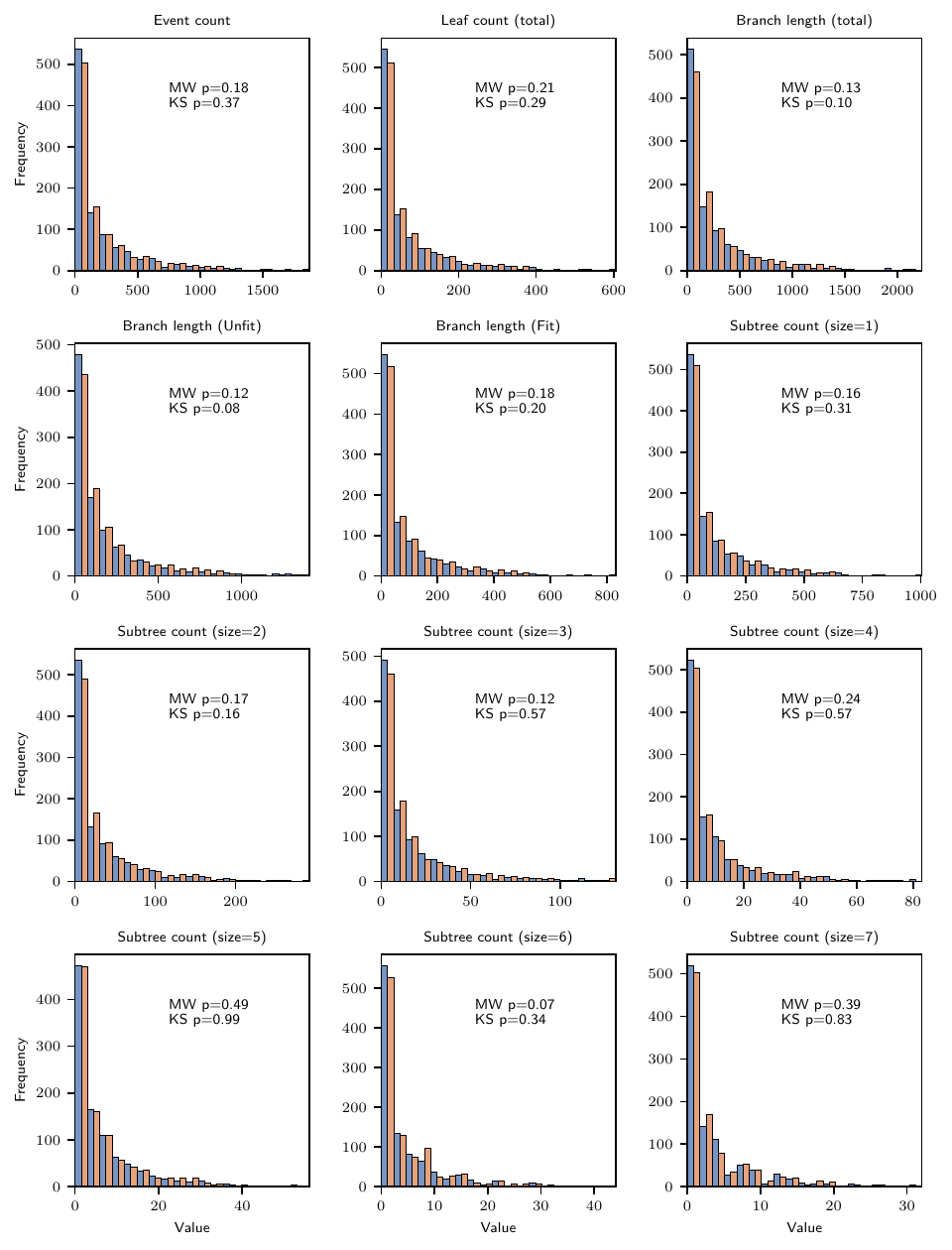}\\
\includegraphics{figures/legend.pdf}
\caption{Distributional comparisons analogous to those appearing in Figure~\ref{fig:dist-test},
for all summary statistics we considered, as described in Section~\ref{sec:dist-test}.
We emphasize that because all of these tests were computed using the same sample of phylogenies, the $p$-values for different tests need not be independent.}
\label{fig:dist-comp}
\end{figure}

\newpage
\setcounter{figure}{1}
\begin{figure}[ht!]
\centering
\includegraphics[width=0.8\textwidth]{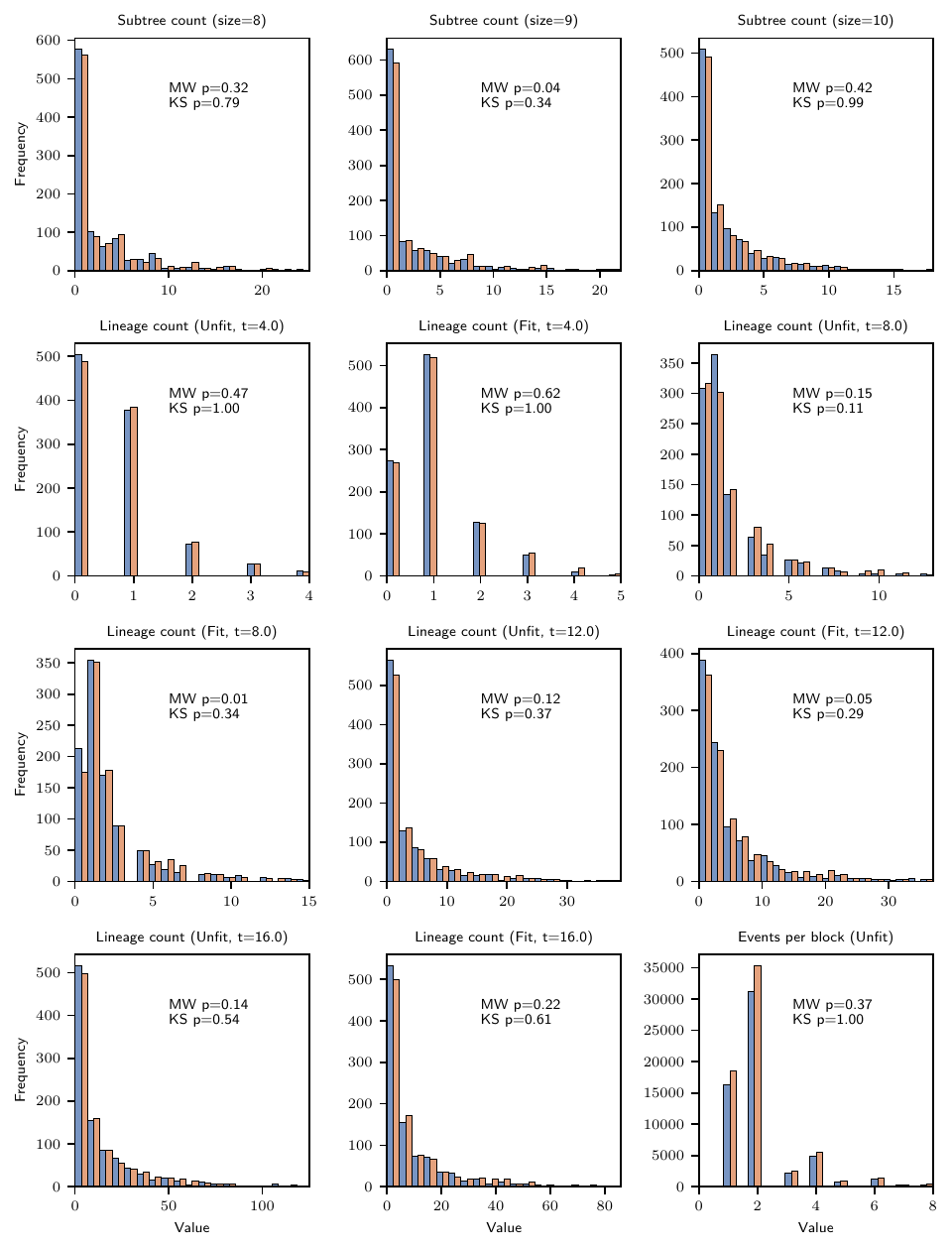}\\
\includegraphics{figures/legend.pdf}
\caption{(Continued)}
\end{figure}

\newpage
\setcounter{figure}{1}
\begin{figure}[ht!]
\centering
\includegraphics[width=0.8\textwidth]{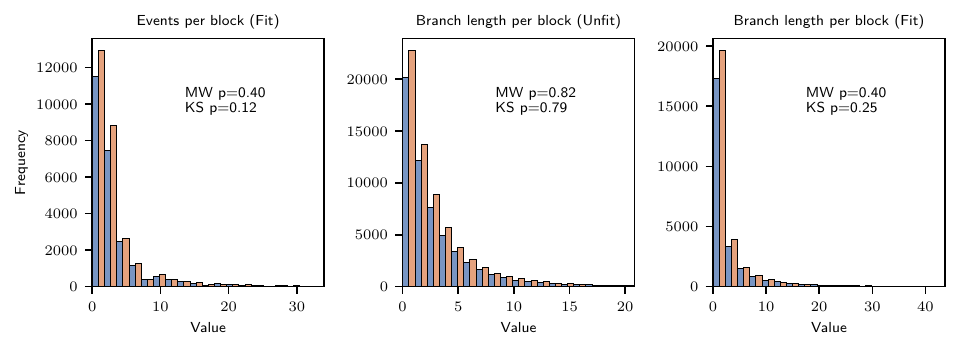}\\
\includegraphics{figures/legend.pdf}
\caption{(Continued)}
\end{figure}

\end{document}